\documentclass[aps,reprint,twocolumns,pre,showpacs,superscriptaddress,floatfix,notitlepage,nofootinbib]{revtex4-1}
\usepackage{amssymb,amsmath,amsfonts} 
\usepackage{amsthm}
\usepackage{times}
\usepackage{mathtools}
\usepackage{physics}
\usepackage{array}
\usepackage{graphicx,epsfig,xcolor}
\usepackage[colorlinks=true,citecolor=blue]{hyperref}
\usepackage{newtxtext,newtxmath}

\begin{document}

\title{Long-range level correlations in quantum systems with finite Hilbert space dimension}

\author{\'{A}ngel L. Corps}
   \email[]{angelo04@ucm.es}
    \affiliation{Departamento de Estructura de la Materia, F\'{i}sica T\'{e}rmica y Electr\'{o}nica \& GISC, Universidad Complutense de Madrid, Av. Complutense s/n, E-28040 Madrid, Spain}
    
\author{Armando Rela\~{n}o}
\email[]{armando.relano@fis.ucm.es}
\affiliation{Departamento de Estructura de la Materia, F\'{i}sica T\'{e}rmica y Electr\'{o}nica \& GISC, Universidad Complutense de Madrid, Av. Complutense s/n, E-28040 Madrid, Spain}

\date{\today} 

\begin{abstract}
We study the spectral statistics of quantum systems with finite Hilbert spaces. We derive a theorem showing that eigenlevels in such systems cannot be globally uncorrelated, even in the case of fully integrable dynamics, as a consequence of the unfolding procedure. We provide an analytic expression for the power-spectrum of the $\delta_n$ statistic for a model of intermediate statistics with level repulsion but independent spacings, and we show both numerically and analytically that the result is spoiled by the unfolding procedure. Then, we provide a simple model to account for this phenomenon, and test it by means of numerics on the disordered XXZ chain, the paradigmatic model of many-body localization, and the rational Gaudin-Richardson model, a prototypical model for quantum integrability.    
\end{abstract}

\maketitle
\section{Introduction}

The last years have witnessed a revival of the interest in quantum chaos and spectral statistics due to the yet to uncover exotic features of many-body quantum systems without a semiclassical analogue \cite{rigol2016}. A closely related concept is that of quantum integrability, which is also
 present in many current
research topics, from non-equilibrium dynamics and thermalization
\cite{Calabrese:16} to many-body localization and condensed matter
\cite{Pepper:17}. 

Normally, integrability is understood as the opposite of chaos. However, its very definition in quantum mechanics
is far from clear \cite{Weigert:92,Caux:11,Yuzbashyan:13}. Classically, a system is integrable if it has as many independent integrals of motion in involution as degrees of freedom \cite{Arnold:78}. However, translating these classical concepts into the quantum realm is impossible altogether as there is no way to define truly independent integrals of motion \cite{Neumann:31}. In passing we note a quantum system is often said to be integrable if it can be solved exactly \cite{Bethe1931,Levkovich2016}.
For this reason the study of level fluctuations has arguably developed into the most common practical tool to identify the signatures of quantum integrability. In this direction, Berry and Tabor proposed in their pioneering work \cite{berrytabor} that level fluctuations of quantum systems whose classical analogue is integrable belong in the universality class of the Poisson point process, and thus they can be described by independent and exponentially distributed level spacings. Generalizing this idea, one of the main signatures of integrability in quantum many-body systems without a clear semiclassical analogue is also the Poissonian and independent character of their level spacings \cite{Caux:11}. Notwithstanding, it is worth noting that the spectral statistics of real quantum systems with an integrable classical limit show well-known deviations from an exact Poissonian behavior \cite{Berry1985}. There are just a small number of analytical results concerning systems with perfectly independent level spacings \cite{Marklof2001}, and numerical experiments show that the spectra of paradigmatic quantum integrable billiards slowly approach sequences of independent spacings as the excitation energy is increased \cite{Robnik1998}.
Anyhow, some authors have also argued that the Berry-Tabor result should also hold outside the semiclassical limit \cite{Enciso2006}. Thus, despite all these facts and the existence of well-known exceptions to this behavior \cite{Benet2003}, among which
we mention quantum systems with just one semiclassical degree of freedom (e.g., the harmonic oscillator) or quantum superintegrable systems \cite{Artemio}, the Poissonian behavior is commonly accepted as a trustworthy signature of integrability in many-body quantum systems (see, e.g.,
\cite{Relano2004}).

Similarly, the chaotic regime cannot be defined either in the quantum world in terms of classical concepts \cite{Gomez11}. Level fluctuations of a quantum system whose classical analogue is completely chaotic follow the expectations of random matrix theory (RMT) \cite{Mehta,bgs}. In this limit, the description is also universal, meaning that the particular features of each Hamiltonian matrix is irrelevant and spectral statistics are dominated by the symmetry class of each system. By extension, a quantum system \textit{without} a semiclassical analogue is \textit{defined} to be chaotic if its level fluctuations can be described by RMT. In this sense, the main feature of chaotic quantum systems is that their spectra exhibit strong level correlations and thus they are qualitatively different from integrable ones. Hence, the transition from integrability to chaos implies the emergence of such correlations at some point. In this work, we deal with a simple model for quantum systems close to integrability that gives rise to spectra composed by independent level spacings \cite{Bogomolny1999,Garcia2006b,Bogomolny2001}. A particular limit of this model leads to a perfectly independent Poissonian sequence of level spacings. For the rest of the cases, it also provides independent level spacings, but showing a certain amount of level repulsion, a trademark of chaotic \cite{bgs} and intermediate (neither fully chaotic, nor integrable) quantum systems \cite{Prosen1993,Brody1973}.

Our main result is that the spectrum of quantum many-body systems with \textit{finite} Hilbert space cannot consist of (globally) independent level spacings. Every spectrum must be transformed onto a sequence of dimensionless energies with average level density equal to one before testing whether its level spacings are independent or not. We show that such a transformation, called {\em unfolding} \cite{Gomez2002}, always introduces spurious long-range correlations if the dimension of the system Hilbert space is finite. Therefore, our result applies to a wide class of quantum many-body systems, covering spin chains and bosonic/fermionic lattice models \cite{Montambaux1993,Hsu1993,Bruus1997}, and notably including those exhibiting many-body localization \cite{Oganesyan2007,Huse2010}.

First, we derive an exact result for the $\delta_n$ statistic \cite{Relano2002}, which accounts for long-range spectral correlations. Then, we perform an extensive numerical test on the disordered XXZ Heisenberg chain \cite{Santos2004}, the prototypical model for the transition to many-body localization, which displays a crossover from chaos to integrability \cite{Serbyn2016}. We find that the behavior of the $\delta_n$ statistic strongly deviates from its expected behavior, even in the integrable limit of the model.
Then, we formulate a theorem showing that the unfolding procedure unavoidably breaks the independence of level spacings, if the dimension of the Hilbert space is finite. Finally, we derive a simple generic unfolding model to account for this effect, and test it by means of the same XXZ Heisenberg chain and a fully integrable Richardson-Gaudin model \cite{Dukelsky2001}.

This paper is organized as follows. In Sec. \ref{definitions} we review the unfolding procedure and define the $\delta_{n}$ statistic as well as the basic quantities that will allow for the study of long-range spectral statistics in our work. In Sec. \ref{model} we derive an analytical result for the $\delta_n$ statistic for a family of intermediate quantum systems with level repulsion and independent level spacings, and we apply it to a physical system exhibiting a crossover from integrability to chaos to show that it provides an incorrect estimate for the level repulsion. In Sec. \ref{breaking} we first argue that this must be a consequence of the unfolding procedure and then show that this is the case for every \textit{finite} quantum system. In Sec. \ref{reunfolding} we introduce a simple model for the unfolding procedure that takes into account the spurious correlations introduced by the unfolding, and then re-derive the result obtained in Sec. \ref{model}; on this occasion, both long and short-range results agree almost perfectly. In Sec. \ref{discussion} we discuss the consequences of this result. Finally, in Sec. \ref{conclusions} we gather the main conclusions of our work. Extensive derivations are deferred to Appendices \ref{prooftheorem1} and \ref{prooftheorem3} for convenience.

\section{Spectral statistics}\label{definitions}

As mentioned in the Introduction, before analyzing level fluctuations a transformation, called \textit{unfolding}, is almost always necessary. Although it can be avoided in the study of short-range spectral statistics (as in the notable case of the adjacent level gap ratio \cite{ratios,Corps2020,Tekur2020,Tekur2018}), long-range spectral statistics, which are precisely the aim of this work, \textit{always} require such a preliminary step. The basic features of this smoothing mechanism are reviewed below. 

  Let $\{E_{i}\}_{i=1}^{N}$ be a sequence of
energies (the eigenvalues of a certain quantum Hamiltonian) in ascending order. The cumulative level density
  function \cite{stockmann}, $\mathcal{N}(E)$, 
counts the number of levels up to energy $E$. One needs to assume that it can be
separated into a smooth part, $\overline{\mathcal{N}}(E)$,
and a fluctuating part, $\widetilde{\mathcal{N}}(E)$, in the form
\begin{equation}
  \label{separation}
  \mathcal{N}(E)=\overline{\mathcal{N}}(E)+\widetilde{\mathcal{N}}(E),
\end{equation}
where
\begin{equation}\label{cumulative}
  \overline{\mathcal{N}}(E):=\int_{-\infty}^{E}\textrm{d}E'\,\overline{\rho}(E').
\end{equation}
Here, $\overline{\rho}(E)$ denotes the smooth part of the
density of states, which varies continuously with $E$. This quantity is
used for the \textit{unfolding transformation}: from the original energies $\{E_{i}\}_{i=1}^{N}$, it provides a new sequence of levels in ascending order $\{\varepsilon_{i}\}_{i=1}^{N}$ as
\begin{equation}
  \label{unfolding_procedure}
  \begin{split}
    E_{i}\mapsto\varepsilon_{i}:=\overline{\mathcal{N}}(E_{i}),\,\,\,\forall i\in\{1,2,\dots, N\}.
  \end{split}
\end{equation}
Ultimately, the unfolding procedure aims to isolate the smooth part of $\mathcal{N}(E)$ from its fluctuating
part \cite{Guhr1998,Gomez2002}. For quantum systems
with clear \textit{classical} analogues, there exists a way to derive an
analytic expression \cite{Gutzwiller1967,Gutzwillerbook,Berry1977},
\begin{equation}
  \label{smoothdensityofstates}
  \overline{\rho}(E)=\frac{1}{(2\pi \hbar)^{D}}\int \textrm{d}\mathbf{p}\textrm{d}\mathbf{q}\,\delta[E-\mathcal{H}(\mathbf{q},\mathbf{p})],
\end{equation}
where $\mathcal{H}(\mathbf{q},\mathbf{p})$ is the classical Hamiltonian, $D\in\mathbb{N}$ is the spatial dimension, and
$(\mathbf{q},\mathbf{p})\in \mathbb{R}^{2D}$ are the usual
position-momentum variables of the phase space. From Eq. \eqref{smoothdensityofstates} it follows that the smooth cumulative level density is a non-negative function for all values of $E$, and thus a correct unfolding procedure yields $\varepsilon_{i}\geq 0$ for all $i\in\{1,\ldots,N\}$ (as opposed to $E_{i}$, which can be any real number).

In this way, we can separate the complete density of states into a smooth part, given by Eq. \eqref{smoothdensityofstates} in the case of semiclassical systems, and a fluctuating part, $\widetilde{\rho}(E):=\rho(E)-\overline{\rho}(E)$. The first one, $\overline{\rho}(E)$, defines the particular
features of each quantum system, whereas the fluctuating part,
$\widetilde{\rho}(E)$, is \textit{universal} as it is associated to integrable or chaotic nature of all quantum systems \cite{Gomez11}. The dimensionless 
\textit{unfolded energies} $\{\varepsilon_{i}\}_{i=1}^{N}$ can be used
to study the \textit{universal} properties of level fluctuations as the smooth, particular features of each system have been eliminated after such a procedure. This is the essence of level fluctuations.

Unfortunately, Eq. \eqref{smoothdensityofstates} can only be solved analytically for few systems and, in any case, strictly speaking, it is only valid for systems having a well-defined semiclassical limit. A generalization
for many-body systems in the mean-field approximation is obtained by
means of the celebrated Bethe formula \cite{Bethe1936,Leboeuf2005}. However, these results are not
directly applicable to {interacting many-body} quantum systems.  Consequently, in the context of quantum many-body systems studying level statistics requires assuming that the separation Eq. \eqref{separation} remains valid; then, one obtains Eqs. \eqref{cumulative} and \eqref{smoothdensityofstates} numerically, a technique widely accepted by the community as the extensive literature reflects \cite{Serbyn2016,Bertrand2016,Torres2017,Santos2004,Corpsb2020,Torres2020,Lerma2019,Schiulaz2019,Luitz2017,Sierant2019PRB,Sierant2020}. In particular, the importance of correctly estimating the smooth part of the level density has been highlighted before \cite{Jackson2001}. Very frequently, no information about the functional form of $\mathcal{\overline{N}}(E)$ is available, and a generic polynomial of a certain degree $\mathcal{\overline{N}}(E)=\sum_{k=0}^{\textrm{deg}}c_{k}E^{k}$ is used to fit the actual cumulative density of the original set of levels $\{E_{i}\}_{i=1}^{N}$. The result of this operation is then used to obtain the unfolded levels $\{\varepsilon_{i}\}_{i=1}^{N}$ as explained above.

One basic quantity to study spectral analysis is the (unfolded) level spacing, $s_{i}:= \varepsilon_{i+1}-\varepsilon_{i}\geq0$, $i\in\{1,\ldots,N-1\}$. The mean value $\langle s_{i}\rangle$ is defined as an average over a ensemble of equivalent spectra. That is, if $s_{i}^{(k)}$ denotes the $i$-th spacing in the $k$-th realization, then on the unfolded scale \begin{equation}\label{unity}
    \langle s_{i}\rangle=\lim_{M\to\infty}\frac{1}{M}\sum_{k=1}^{M}s_{i}^{(k)}=1,\,\,\,\forall i\in\{1,\dots, N\}.
\end{equation}
As this equation is actually independent on $i$, one usually writes $\langle s\rangle := \langle s_{i}\rangle$, $\forall i$. Assuming (statistical) ergodicity, one has the standard ensemble mean $\langle s\rangle= \lim_{N\to\infty}\frac{1}{N}\sum_{i=1}^{N}s_{i}$. It is important to observe that the sample estimator of the mean $\langle s\rangle_{N}$ and the ensemble mean $\langle s\rangle$ may in principle differ somewhat, and thus, clearly,
\begin{equation}
    \langle s\rangle_{N}:=\frac{1}{N}\sum_{i=1}^{N}s_{i}\neq\langle s\rangle=\lim_{N\to\infty}\langle s\rangle_{N}.
\end{equation}
 As a consequence, on the unfolded scale $\langle\varepsilon_{n}\rangle=n\langle s\rangle$. From the spacings one may obtain the celebrated nearest-neighbor spacing distribution (NNSD), which measures short-range level fluctuations between adjacent levels, $P(s):=\langle \delta(s-s_{i})\rangle$.

 In this work we will analyze long-range spectral correlations by means of the $\delta_{n}$ statistic \cite{Relano2002}, which was conceived by drawing an analogy with a discrete time series.  This quantity represents the deviation of the excitation energy of the $(n+1)$-th unfolded level from its mean value in an equiespaced spectrum, $\langle \varepsilon_{n}\rangle=n$:
\begin{equation}\label{deltan}\delta_{n}:= \sum_{i=1}
^{n}\left(s_{i}-\langle s\rangle\right)=\varepsilon_{n+1}-\varepsilon_{1}-n,\end{equation} 
for all $n\in\{1,2,\dots,N-1\}$. One can take $n$ to be a discrete-time index so that the string $\{\varepsilon_{i}\}_{i=1}^{N}$ of length $N\gg 1$ represents a random process.  Then, a discrete Fourier transform can be applied to the statistic. The quantity of interest for the statistical analysis of long-range spectral correlations is actually the (averaged) \textit{power-spectrum}  \cite{Relano2002,demo,Faleiro2006,Gomez2005,Pachon2018,Relano2008,Relano2008b, Santhanam2005,Santhanam2006,Garcia2006} of Eq. \eqref{deltan}:
\begin{equation}\label{powerspectrum}
\langle P_{k}^{\delta}\rangle:=\langle |\mathcal{F}(\delta_{n})|^{2}\rangle=\left\langle\left|\frac{1}{\sqrt{N}}\sum_{n=1}^{N}\delta_{n}e^{-i\omega_{k}n}\right|^{2} \right\rangle,\end{equation}
Here, $\{\omega_{k}\}_{k=1}^{N-1}$ is a set of \textit{dimensionless frequencies} given by $\omega_{k}:=2\pi k/N$ for each $k\in\{1,2,\ldots,N-1\}$. Main results will be plotted up to the Nyquist frequency, $k_{\textrm{Ny}}:=N/2$, but are valid throughout the entire range of frequencies.  In the domain $0<\omega_{k}\ll1$, in quantum integrable systems $\langle P_{k}^{\delta}\rangle$ exhibits the neat power-law decay $\langle P_{k}^{\delta}\rangle\simeq 1/\omega_{k}^{2}$, whereas for quantum chaotic ones this is $\langle P_{k}^{\delta}\rangle\simeq 1/\omega_{k}$ \cite{Relano2002}. This feature is universal inasmuch as it merely depends on the regularity class (integrable or chaotic) of the system but not on its particular symmetries. In the case of semiclassical systems, Eq. \eqref{powerspectrum} has been used to identify non-universal features due, e.g., to short periodic orbits \cite{Dietz17graphs, Faleiro2006}. 

\section{A class of quantum systems with independent spacings}\label{model}

\subsection{The model}

As advanced in the Introduction, we focus on a model which generates independent level spacings, and gives rise to a Poissonian sequence in the appropriate limit. This is the famous
short-range plasma model introduced by Bogomolny and co-workers in Ref. \cite{Bogomolny1999}. It has been particularly successful, e.g., in the study of the metal-insulator transition in the Anderson model \cite{Shklovskii1993,Guhr1998}, where a universal statistics called semi-Poisson appears at the mobility edge.  This kind of intermediate statistics is also present in nonrandom Hamiltonians with a step-like singularity \cite{Garcia2006b}, as well as in a Coulomb billiard \cite{Altshuler1997}, anisotropic Kepler problems \cite{Wintgen1988}, generalized kicked rotors \cite{Hu1999}, pseudointegrable billiards \cite{Bogomolny1999,Bogomolny2001} and others \cite{Bogomolny2004}. Within the context of many-body quantum systems, several variations of this short-range plasma model and further generalizations have been put forward to describe the level statistics of the region between the ergodic (chaotic) and localized (integrable) phases in the many-body localization transition \cite{Serbyn2016,Sierant2020,SierantPRB2020}. 

The short-range plasma model has a joint distribution of eigenvalues equivalent to that of a one-dimensional classical Coulomb gas with $N+2$ particles at equilibrium positions $\{x_{i}\}_{i=0}^{N+1}$ in an interval of length $I$ interacting through a pairwise repulsive logarithmic potential restricted to a finite number of neighbors, $0<j-i\leq h$. If we only consider nearest-neighbor interactions, $h=1$, this is $V(x_{0},x_{1},\ldots,x_{N+1})=-\sum_{i}\log(x_{i}-x_{i-1})$ together with the boundary condition $0=x_{0}<x_{1}<\ldots<x_{N}<x_{N+1}=I$. Under these circumstances, in the large $N$ limit the corresponding NNSD can be shown to yield 
\begin{equation}\label{psgeneralized}
P(s;\eta):=\frac{\eta^{\eta}s^{\eta-1}e^{-\eta s}}{\Gamma(\eta)},\hspace{0.5cm}s\geq0,\,\,\,\eta\in[1,+\infty),
\end{equation}
where $\Gamma(\eta):=\int_{0}^{\infty}\textrm{d}t\,t^{\eta-1}e^{-t}$. Eq. \eqref{psgeneralized} is Eq. (5) in Ref. \cite{Bogomolny1999} with $\eta:=\beta+1$ (and $n=1$), and it correctly reproduces known results such as the Poissonian case, $P(s;\eta=1)=e^{-s}$ or the semi-Poisson $P(s;\eta=2)=4se^{-2s}$. We note that although the semi-Poissonian limit strictly corresponds to $\eta=2$, the term is sometimes used to mean the entire family of distributions Eq. \eqref{psgeneralized}. When $\eta$ departs from the fully Poissonian limit ($\eta=1$), the NNSD reveals level repulsion of the form $P(s)\propto s^{\eta-1}$, but asymptotically it decreases much more slowly than in the case of quantum chaotic systems, for which the Wigner-Dyson surmise $P_{\textrm{WD}}(s):=a_{\beta}s^{\beta}e^{-b_{\beta}s^{2}}$ applies [where $\beta\in\{1,2,4\}$ here is the usual Dyson symmetry index] \cite{stockmann}.  
The NNSD of the classical random matrix ensembles describing fully chaotic spectra \cite{Mehta,bgs} are \textit{not} included in this formula as spacing correlations are absent from the model.

It is noteworthy that the underlying short-range plasma model can be regarded as an ensemble of free particles in a bath at finite temperature, $T=1/\beta$ (to make contact with statistical mechanics, we note that the Boltzmann constant has been set $k_{B}=1$ here). 
The inverse temperature $\beta\in[0,+\infty)$ can be understood as a \textit{continuous} repulsion index of random matrix ensembles: for small $\beta$, the bath temperature $T$ is large, thermal energy wins over the logarithmic potential interaction, and the gas particles move away significantly from their equilibrium positions; this is equivalent to level clustering in regular spectra \cite{berrytabor}, where levels can be degenerate and thus potentially overlap. By contrast, large $\beta$ implies little displacement of the gas particles from equilibrium and a spectrum where level repulsion is a relevant effect, mimicking avoided level crossings and spectral rigidity in chaotic spectra.

It has been shown \cite{Saldana1999,Garcia2006b} that the level fluctuations of this model are equivalent to those of a spectrum obtained by
keeping every $\eta\in\mathbb{N}$ eigenvalues from an initial independent Poissonian spectrum (to highlight this interpretation we have chosen $\eta$ instead of the usual Dyson index $\beta$), in a daisy-like model fashion. Doing so, one gets a new sequence of spacings $\{\overline{s}_{i}\}_{i}$ where \begin{equation}\label{spacingbarra}
\overline{s}_{i}:=\frac{1}{\eta}\sum_{j=0}^{\eta-1}s_{i+j}=\frac{1}{\eta}(s_{i}+s_{i+1}+\ldots+s_{i+\eta-1}),\,\,\,i=1,1+\eta,1+2\eta,\ldots
\end{equation}
For simplicity, in what follows the spacings defined in Eq. \eqref{spacingbarra} will still be denoted by $s_{i}$. It can be shown that the sum of $\eta$ independent Poissonian random variables (in this context, exponential random variables with mean $\lambda=1$) is an \textit{Erlang distribution} with shape parameter $\eta\in\mathbb{N}$ (and rate $\lambda=1$), whose probability density function is $\mathcal{P}(s)=s^{\eta-1}\exp(-s)/(\eta-1)!$
This can be further generalized by regarding the distribution $\mathcal{P}(s)$ as a function of the parameter $\eta$ itself, $\mathcal{P}(s;\eta)$. In that case, we can analytically  extend its domain, $\eta\in\mathbb{N}\to[1,+\infty)$. Then, Eq. \eqref{psgeneralized} is completely equivalent \cite{exp1} to the marginal probability density of any given spacing Eq. \eqref{spacingbarra}, i.e., $P(s;\eta)=\eta\,\mathcal{P}(\eta s;\eta)$.

Due to the independence of the level spacings, it is easy to show that if the set  $\{s_{i}\}_{i=1}^{N}$ are distributed as in Eq. \eqref{psgeneralized}, then 

\begin{equation}\label{sisj}
    \langle s_{i}\rangle=1,\hspace{1cm}\,\,\,\,\langle s_{i}s_{j}\rangle=1+\frac{\delta_{ij}}{\eta},
\end{equation}
for all $i,j\in\{1,2,\ldots,N\}$. Here, $\delta_{ij}$ is the Kronecker delta. Thus, in particular, 

\begin{equation}\label{eqcovar}
\textrm{Cov}(s_{i},s_{j})=\langle s_{i}s_{j}\rangle-\langle s_{i}\rangle \langle s_{j}\rangle=\delta_{ij}/\eta,
\end{equation}
which vanishes unless $i=j$. Thus we observe that the spacings of the family of distributions in Eq. \eqref{psgeneralized} show properties that are intermediate between those of quantum chaotic (level repulsion) and integrable (statistical independence) systems. 

\subsection{Exact result for the $\delta_n$ statistic}

Some results concerning the long-range spectral statistics for this model are well-known and were derived some time ago \cite{Bogomolny1999,Bogomolny2001}. The asymptotic behavior of one of the most used statistics, the number variance, $\Sigma^2(L)$, is quite simple, $\Sigma^2(L) \sim L/\eta$ ($L\to\infty$); however, its exact analytical expression is highly involved, even for integer values of the parameter $\eta$. Here, we provide an \textit{exact} and very simple expression for the (averaged) power-spectrum $\langle P_{k}^{\delta}\rangle$, valid for any value of $\eta \geq 1$.

\textbf{Theorem 1.} Let $\{\varepsilon_{i}\}_{i=1}^{N+1}$ be an (unfolded) finite quantum spectrum of $N+1$ levels \cite{exp2} giving rise to the $N$ independent and identically distributed set of spacings $\{s_{i}\}_{i=1}^{N}$ where each spacing follows the distribution Eq. \eqref{psgeneralized}. Then, the power-spectrum of the $\delta_{n}$ statistic can be written 
\begin{equation}\label{powerspectrumnounfolding}
\langle P_{k}^{\delta}\rangle=\frac{1}{2\eta\,\sin^{2}(\omega_{k}/2)},\end{equation}
where $k\in\{1,2,\ldots,N\}$ and $\eta\in[1,+\infty)$. 

\textit{Proof.} See Appendix \ref{prooftheorem1}.

The above formula is completely general and remains valid for any $\eta\in[1,+\infty)$ taken as a continuous parameter. We note that a mathematically equivalent formula was previously found without considering the explicit underlying model of intermediate statistics Eq. \eqref{psgeneralized} (but also with uncorrelated spacings) \cite{Riser}. 

\subsection{Numerical test}

It is not easy to find physical systems well described by the short-range plasma model discussed above. Here, we rely on one of the paradigmatic models in studies of many-body localization, which is approximately described by several short-range plasma models \cite{Serbyn2016,Sierant2020,SierantPRB2020}, and includes an integrable limit. This is the XXZ Heisenberg chain \cite{Serbyn2016,Bertrand2016,Torres2017,Santos2004,Corpsb2020,Luitz2017}. Its Hamiltonian can be written as
\begin{equation}
  \label{modelxxz}
  \begin{split}
  \mathcal{H}_\textrm{XXZ}&:=\sum_{\ell=1}^{L}\omega_{\ell}\hat{S}_{\ell}^{z}+J\sum_{\ell=1}^{L-1}\left(\hat{S}_{\ell}^{x}\hat{S}_{\ell+1}^{x}+\hat{S}_{\ell}^{y}\hat{S}_{\ell+1}^{y}+
  \hat{S}_{\ell}^{z}\hat{S}_{\ell+1}^{z}\right),
\end{split}\end{equation}
which is a one-dimensional chain with two-body nearest-neighbor
couplings, $L$ sites, and random onsite magnetic fields, $\omega_{\ell}$. Here, $\hat{S}_{\ell}^{x,y,z}$ are the total spin operators at site $\ell\in\{1,\ldots,L\}$.
We choose $J=1$ and have defined $\hbar:= 1$. The difference with its clean analogue is that we introduce disorder by means of uniformly, randomly distributed magnetic fields $\omega_{\ell}\sim \mathcal{U}(-\omega,\omega)$. In the clean limit, $\omega=0$, the system is integrable and can be described by the Bethe-ansatz \cite{Nandkishore2015,rigol2016}. For intermediate values of $\omega$, the chain exhibits a chaotic phase where spectral statistics very approximately coincide with those of RMT. For $L=14$, this region comprises the disorder strength range $0.3\lesssim \omega \lesssim 1.4$ \cite{Corpsb2020}.  Finally, for $\omega\gg 1$ the model enters the many-body localized (MBL) phase, characterized by integrable dynamics \cite{Serbyn2013,Ros2015}. Thus, level statistics of both the Bethe-ansatz and the MBL phases obey the Poissonian limit.

The community of many-body localization commonly considers the eigenvalues associated to the eigenstates of $\hat{S}^{z}:=\sum_{i}\hat{S}_{i}^{z}$, which commutes with the Hamiltonian, $[\mathcal{H},\hat{S}^{z}]=0$. Thus, we can restrict ourselves to the sector $S^{z}=0$, where $S^{z}$ is the eigenvalue of the operator $\hat{S}^{z}$. The dimension of the Hilbert space is then $d=\binom{L}{L/2}$. We consider $L=14$, so $d=3432$. The semiclassical limit of this system is obtained in the limit of large spin size. Since here we are concerned with a chain of $1/2$ spins, this situation cannot be reached, not even when the number of sites $L\to\infty$. Thus, as there exists no underlying statistical theory that provides $\overline{\rho}(E)$ for this model, to unfold we have performed a numerical fit to the energies with a polynomial of degree 6. Only the central $N_{\textrm{unf}}=d/3=1144$ have been used to this end, and the spectral statistics have been analyzed with the central $N=d/4=864$ levels after unfolding. We have averaged over 1000 realizations for each value of the random disorder $\omega$.

\begin{center}
\begin{figure}
\includegraphics[width=0.49\textwidth]{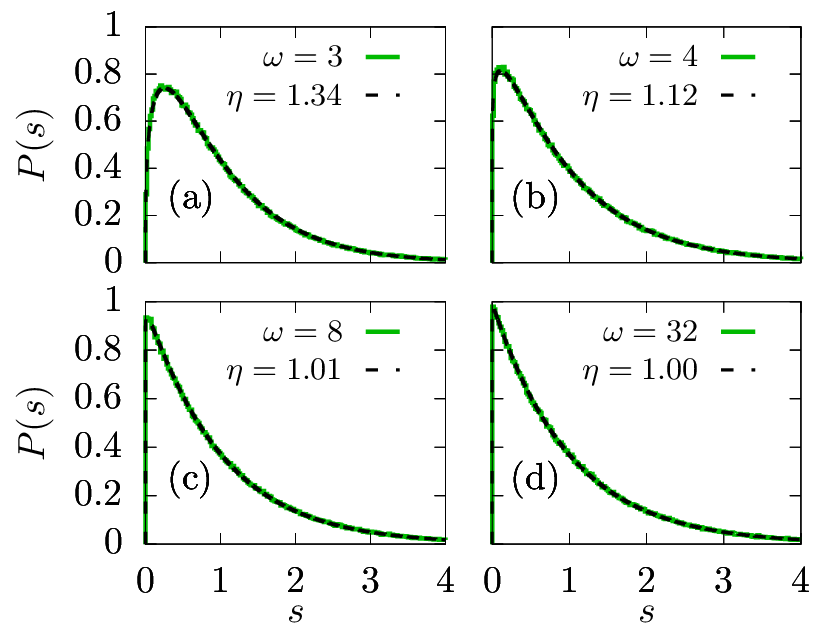} 
\caption{Nearest-neighbor spacing distribution, $P(s)$, for the disordered XXZ Heisenberg chain, Eq. \eqref{modelxxz}. The number of sites is $L=14$. The black, dashed line corresponds to best fit of the model of intermediate statistics Eq. \eqref{psgeneralized} to the numerically obtained $P(s)$. As explained in the text, $N/N_{\textrm{unf}}=0.75$. Panels (a)-(d) show the results for disorder strengths $\omega$ and the corresponding value of the repulsion parameter $\eta$ obtained from the fit.  }
\label{psxxz}
\end{figure}
\end{center}

\begin{center}
\begin{figure}
\includegraphics[width=0.49\textwidth]{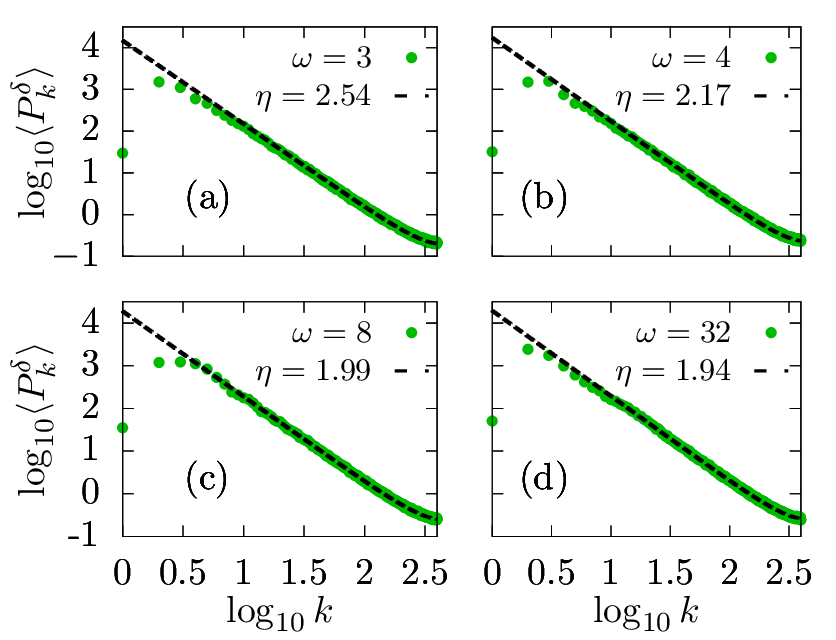} 
\caption{Averaged $\delta_{n}$ power-spectrum, $\langle P_{k}^{\delta}\rangle$ for the disordered XXZ Heisenberg chain, Eq. \eqref{modelxxz}. The number of sites is $L=14$. The black, dashed line corresponds to best fit of the model of intermediate statistics Eq. \eqref{powerspectrumnounfolding} to the numerically obtained $\langle P_{k}^{\delta}\rangle$. As explained in the text, $N/N_{\textrm{unf}}=0.75$. Panels (a)-(d) show the results for disorder strengths $\omega$ and the corresponding value of the repulsion parameter $\eta$ obtained from the fit.}
\label{powxxz}
\end{figure}
\end{center}

The results for the NNSD are shown in Fig. \ref{psxxz}, while those for the $\delta_{n}$ power-spectrum can be found in Fig. \ref{powxxz}. As can be seen, the fit to Eqs. \eqref{psgeneralized} and \eqref{powerspectrumnounfolding} is almost perfect in all cases. At first sight one might conclude that the short-range plasma model provides a good description of the spectral statistics of the XXZ Heisenberg chain when it is far away from the ergodic regime, and this conclusion should be compatible with more sophisticated analysis involving a broader class of plasma models \cite{Serbyn2016,Sierant2020,SierantPRB2020}. However, we must draw attention to the extracted value of the repulsion parameter $\eta$ of Eqs. \eqref{psgeneralized} and \eqref{powerspectrumnounfolding}. Strikingly, they are \textit{different} by a factor of approximately 2. Therefore, and in contrast to our previous na\"{\i}ve statement, these results and theorem 1 allow us to conclude that the spectrum of the XXZ Heisenberg chain cannot be composed of (globally) independent level spacings, not even in the integrable limit reached at very large values of $\omega$, as long-range spectral correlations (the power-spectrum $\langle P_{k}^{\delta}\rangle$) do not reproduce the corresponding short-range result (the NNSD) based on statistical independence, not even approximately.

There are a number of possible explanations for this fact. One of them is the existence of an effect similar to the Berry's saturation \cite{Berry1985} in systems with no semiclassical limit. However, such an effect seems difficult to justify, due to the \textit{absence of a clear semiclassical model} and their associated periodic orbits. Hence, we focus on another possible explanation: \textit{the consequences of the unfolding procedure}. In the next two sections we formulate two theorems. The first one states that this preliminary step implies that the nearest neighbor spacings cannot be (globally) uncorrelated in \textit{finite} quantum systems. The second explains the factor of 2 linking Figs. \ref{psxxz} and \ref{powxxz} relying on a simple model that incorporates the spurious effects \cite{Gomez2002} of unfolding.

\section{Breaking of spacings independence}\label{breaking}

The $\delta_n$ statistic, Eq. \eqref{deltan}, provides a simple picture for long-range spectral statistics. If spacings are independent, $\delta_n$ behaves like a \textit{random walk} \cite{Relano2002}. We can interpret each spectrum as a random walker starting from `home', $\delta_0 \equiv 0$, and advancing a random distance at each step $n$, with no memory of its previous path. In this section we prove that the unfolding procedure implies that $\delta_{N-1} \leq 1$, being $N$ the size of the Hilbert space \cite{exp3}. This means that every walker must be close to `home' after its last step, $n=N-1$. This is only possible if, at some point along its path, every walker decides to come back home, and arranges its wandering in such a way as to reach this goal. More precisely, {\em the unfolding procedure means that energy levels of every finite quantum system must be somehow correlated, regardless of whether it is integrable or not.}

To formulate this theorem, we make two assumptions. The first one is that the smooth part of the cumulative level density, $\mathcal{\overline{N}}(E)$, must be a non-decreasing function of energy. To see why this requirement must be fulfilled by any cumulative level density, let us imagine a quantum system in which we can find two consecutive eigenlevels, $E_{i}$ and $E_{i+1}$ such that $E_{i+1} > E_i$ and $\mathcal{\overline{N}}(E_{i+1}) < \mathcal{\overline{N}}(E_i)$. In such a case, Eq. \eqref{unfolding_procedure} establishes that the corresponding unfolded levels fulfill $\varepsilon_{i+1} < \varepsilon_i$, and therefore the corresponding level spacing, $s_i = \varepsilon_{i+1} - \varepsilon_i$ is negative. Therefore, as the nearest neighbor level spacings must always be non-negative, we conclude that this assumption is not only reasonable, but mandatory to perform a proper spectral fluctuation analysis. 

Our second assumption is that the smooth part of the level density, $\overline{\rho}(E)$ is normalized to the dimension of the Hilbert space, $N$, which means that $\lim_{E \rightarrow -\infty} \mathcal{\overline{N}}(E)=0$ and $\lim_{E \rightarrow \infty} \mathcal{\overline{N}}(E)=N$. This is a very reasonable assumption which holds under very different circumstances. For example, in many-body quantum systems with a well-defined semiclassical analogue, Eq. \eqref{smoothdensityofstates} is normalized to the total number of energy levels, $N$; this happens, e.g., in the Lipkin-Meshkov-Glick model \cite{LMG}. The same normalization holds for standard random matrix ensembles, giving rise to Wigner's semicircular law for the smooth part of its density of states \cite{Mehta}, and embedded random matrix ensembles \cite{Kotabook}, well described by a smooth Gaussian density of states \cite{Gomez11}. A more general reasoning can be argued as follows. By definition, the full cumulative level density verifies $\lim_{E \rightarrow -\infty} \mathcal{N}(E)=0$ and $\lim_{E \rightarrow \infty} \mathcal{N}(E)=N$, if the dimension of the corresponding Hilbert space is $N<\infty$. Now, let us assume that a particular quantum system belonging to this class is described by means of a function $\mathcal{\overline{N}}(E)$ which verifies $\lim_{E \to \infty} \mathcal{\overline{N}}(E)=M>N$. In such a case, all the energy levels above a certain threshold, $E_j$, verify $\mathcal{\overline{N}}(E_i)>N$, $\forall \; i \geq j$. And therefore, the fluctuating part of the cumulative density of states, $\mathcal{\widetilde{N}}(E) = \mathcal{N}(E) - \mathcal{\overline{N}}(E)$, for all these states verify $\mathcal{\widetilde{N}}(E_i)<0$, $\forall \; i \geq j$. This would mean that the fluctuating part of the cumulative density of states would not fluctuate around zero, as expected; contrarily, it would display a systematic and \textit{permanent} negative trend. Since the same conclusions are obtained if $M<N$, and if applied to the lower bound, we have that this second assumption is also mandatory to perform a good unfolding procedure.

It is worth noting that these two assumptions not only mean that $\lim_{E\to-\infty}\mathcal{\overline{N}}(E)=0$ and $\lim_{E\to\infty}\mathcal{\overline{N}}(E)=N$, but also imply that $0 \leq\mathcal{\overline{N}}(E)\leq N$ for \textit{any} value of $E$. 
Then, the maximum value that any unfolded level can take, $\max \varepsilon_{n}=\max \mathcal{\overline{N}}(E_{n})$, is, from the previous discussion, $\max \varepsilon_{n}=N$, and correspondingly the minimum value is $\min \varepsilon_{n}=0$, for all $n\in\{1,\ldots,N\}$

From all these facts, we formulate and prove the following theorem:

\textbf{Theorem 2.} Let $1\ll N<\infty$ be the dimension of any quantum system with a \textit{finite} Hilbert space with eigenenergies $\{E_{i}\}_{i=1}^{N}$. Suppose that the smooth cumulative level density verifies (i) $\mathcal{\overline{N}}(E)$ is a non-decreasing function of $E$, (ii) $\lim_{E\to-\infty}\mathcal{\overline{N}}(E)=0$, and (iii) $\lim_{E\to\infty}\mathcal{\overline{N}}(E)=N$. Then, the $\delta_{n}$ statistic evaluated at $n=N-1$ verifies

\begin{equation}\label{deltanmenor1}
    \delta_{N-1}\leq 1.
\end{equation}

\textit{Proof.} Starting from the spectrum $\{E_{i}\}_{i=1}^{N}$, we unfold by separating the density of states in its smooth and fluctuating parts according to Eq. \eqref{unfolding_procedure}, that is,
the $n$-th unfolded level $\varepsilon_{n}$ is  $\varepsilon_{n}=\mathcal{ \overline{N}}(E_n)$. 
From the definition of $\delta_{n}$, Eq. \eqref{deltan}, we have that at $n=N-1$ \begin{equation}
    \delta_{N-1}=\varepsilon_{N}-\varepsilon_{1}-(N-1)=(\varepsilon_{N}+1)-(\varepsilon_{1}+N).
\end{equation}
Since all the unfolded levels, $\left\{ \varepsilon_i \right\}_{i=1}^N$, are ordered, $\varepsilon_1 \leq \cdots \leq \varepsilon_N$, the maximum of this quantity is obtained by maximizing the first parenthesis and minimizing the second one. As argued before, the conditions above imply that $\mathcal{\overline{N}}(E)\leq N$ for all $E\in\mathbb{R}$, and thus the maximum of $\varepsilon_{N}$ is $\max \varepsilon_{N}=N$. On the other hand, as $\mathcal{\overline{N}}(E)\geq 0$, the minimum of $\varepsilon_{1}$ is $\min\varepsilon_{1}=0$. This means that
\begin{equation}
    \delta_{N-1}\leq (N+1)-(0+N)=1.
\end{equation}
This proves Eq. \eqref{deltanmenor1}. $\hfill\blacksquare$

From the previous result, we reach the following conclusion: 

\textbf{Corollary.} The spectrum of \textit{finite} quantum systems cannot be (globally) composed of statistically independent spacings.

\textit{Proof.} We first note that, by definition, $\delta_{n=N-1}$ oscillates around zero, i.e., \begin{equation}
    \langle\delta_{N-1}\rangle=\langle\varepsilon_{N}\rangle-\langle\varepsilon_{1}\rangle-(N-1)=0.
\end{equation}
For a set of \textit{fully} statistically independent spacings [which in this case in particular follow Eq. \eqref{psgeneralized}], the correlator $\langle\delta_{\ell}\delta_{m}\rangle$ is given by Eq. \eqref{a5}. In conjunction with \eqref{eq11}, this implies that the variance of $\delta_{N-1}$ is \begin{equation}
    \textrm{Var}[\delta_{N-1}]=\langle \delta_{N-1}^{2}\rangle-\langle\delta_{N-1}\rangle^{2}=\frac{N-1}{\eta}.
\end{equation} 
By the central limit theorem, for asymptotically large values of $N$ the distribution of $\delta_{N-1}$ approaches a Gaussian of mean $0$ and variance $\textrm{Var}[\delta_{N-1}]\propto N-1$. This is incompatible with Eq. \eqref{deltanmenor1}. Therefore, it is impossible for the spacings to be (globally) independent. $\hfill\blacksquare$

It is noteworthy that the same argument used to prove Theorem 2 implies that the unfolding procedure establishes a bound for the $\delta_n$ statistic at every value of $n$, $\delta_n \leq N-n$. Therefore, as by definition the $\delta_n$ statistic is not bounded from above for spectra composed by independent spacings with NNSD given by Eq. \eqref{psgeneralized}, we can conclude that the unfolding procedure globally spoils the expected behavior for the $\delta_n$ statistic. Notwithstanding, as $\textrm{Var}[\delta_n] \propto n$, this bound only becomes important when the value of $n$ is close to the dimension of the Hilbert space, $N$. We will discuss this point in detail in Sec. \ref{discussion}.

\section{A simple model for the unfolding procedure}\label{reunfolding}

From Eq. \eqref{deltan}, it is easy to see that Theorem 2 also establishes an upper bound for the mean level spacing,
\begin{equation}
\label{bound_for_s}
    \langle s\rangle_{N}=\frac{1}{N} \sum_{i=1}^{N}s_{i} \leq 1 + \frac{1}{N}.
\end{equation}
The aim of this section is to propose a simple model to account for the consequences of this fact. Eq. \eqref{bound_for_s} entails that, due to the unfolding procedure, the sample estimator of the mean of the level spacing is always very close to the ensemble mean, $\left< s \right>_{N} \approx \left<s\right>=1$ (as usually $N\gg 1$). Hence, we derive and prove a theorem that provides an exact expression for the power-spectrum of the $\delta_n$ statistic if $\left< s \right>_{N}= \left<s\right>=1$ exactly. Given Eq. \eqref{bound_for_s}, it is reasonable to assume that this theorem will provide very accurate results for quantum systems with finite Hilbert spaces of dimension $1\ll N<\infty$, for which Theorem 2 holds. 

\subsection{Main result}
\textbf{Theorem 3.} Let $\{\varepsilon_{i}\}_{i=1}^{N+1}$ be an (unfolded) finite quantum spectrum of $N+1$ levels giving rise to the $N$ independent and identically distributed set of spacings $\{s_{i}\}_{i=1}^{N}$, where each spacing is distributed as in Eq. \eqref{psgeneralized}. Suppose that the unfolding procedure allows to obtain the new \textit{re-unfolded spacings} $\{\widetilde{s}_{i}\}_{i=1}^{N}$, where the $i$-th spacing is defined \begin{equation}\label{scorr}
\widetilde{s}_{i}:=\frac{s_{i}}{\langle s_{}\rangle_{N}}=\frac{Ns_{i}}{\sum_{k=1}^{N}s_{k}},\,\,\forall i\in\{1,\dots,N\}. 
\end{equation} Then, the power-spectrum of the $\delta_{n}$ spectral statistic is \begin{equation}\label{power4}\langle {P}_{k}^{\delta}\rangle=\left(\frac{N}{\eta N+1}\right)\frac{1}{4\sin^{2}(\omega_{k}/2)},
\end{equation}
where $k\in\{1,2,\ldots,N\}$ and $\eta\in[1,+\infty)$. 

\textit{Proof.} See Appendix \ref{prooftheorem3}. 

We first draw attention to the fact that for large enough sequences of spacings, $1\ll N<\infty$, the above result Eq. \eqref{power4} is essentially Eq. \eqref{powerspectrumnounfolding} with $\eta\mapsto 2\eta$, which provides a hint about the disagreement between Figs. \ref{psxxz} and \ref{powxxz}. In the domain $\omega_{k}\ll 1$, one finds the known inverse square power-law $1/f^{2}$, characteristic of fully integrable spectra \cite{Relano2002}. However, now we observe that this feature is actually preserved as long as the \textit{initial} set of spacings are uncorrelated as in Eqs. \eqref{spacingbarra} and \eqref{psgeneralized} (but not necessarily Poissonian): \begin{equation}\langle P_{k}^{\delta}\rangle\simeq \frac{N}{\eta N+1}\left[\frac{1}{\omega_{k}^{2}}+\frac{1}{12}+\mathcal{O}(\omega_{k}^{2})\right],\,\,\,0<\omega_{k}\ll 1. \end{equation} To lowest order in $\omega_{k}$ this reads $
\langle P_{k}^{\delta}\rangle\simeq 1/{\omega_{k}^{2}}\propto 1/{k^{2}}$. The same behavior can be read off Eq. \eqref{powerspectrumnounfolding}.

It is worth to remark that the set of \textit{re-unfolded spacings} of the previous theorem, $\{\widetilde{s}_{i}\}_{i=1}^{N}$, are correlated (even though the initial spacings $\{s_{i}\}_{i=1}^{N}$ are not). To see this, we can calculate the covariance of any two spacings $\widetilde{s}_{i}$ and $\widetilde{s}_{j}$ [see Eqs. \eqref{mean1} and \eqref{a15} of Appendix \ref{prooftheorem3}], which gives
\begin{equation}\label{cova}\begin{split}
\textrm{Cov}(\widetilde{s}_{i},\widetilde{s}_{j})=\langle \widetilde{s}_{i}\widetilde{s}_{j}\rangle-\langle\widetilde{s}_{i}\rangle\langle\widetilde{s}_{j}\rangle=\frac{N(\delta_{ij}+\eta)}{\eta N+1}-1,
\end{split}\end{equation}
for all $i,j\in\{1,\dots,N\}$. Eq. \eqref{cova} is nonzero except if $i=j$ and $N=1$, corresponding to the trivial case where $\{\widetilde{s}_{i}\}_{i=1}^{N}=\widetilde{s}_{1}\equiv 1$. Thus, the sequence of random variables $\{\widetilde{s}_{i}\}_{i=1}^{N}$ are correlated. 

\subsection{Numerical test}

We begin with a simple-minded, pedagogical example to illustrate the forthcoming results involving actual physical systems.  We consider a set of $N<\infty$ statistically independent numbers $\{E_{i}\}_{i=1}^{N}$, where each $E_{i}$ is a Gaussian number of mean $0$ and variance $1$, $E_{i}\sim \mathcal{G}(0,1)$. This set gives rise to an \textit{ad hoc} integrable system, whose spectrum is obtained simply by ordering this sequence of numbers in ascending order. By definition, the smooth part of its cumulative level function is     
\begin{equation}\label{accnormal}\begin{split}
    \mathcal{\overline{N}}(E)&=N\int_{-\infty}^{E}\mathrm{d}x\,\frac{1}{\sqrt{2\pi}}\exp\left(\frac{-x^{2}}{{2}}\right) =\frac{N}{2}\left[1+\erf\left(\frac{E}{\sqrt{2}}\right)\right],
\end{split}\end{equation}
where $\erf(x):=2\int_{0}^{x}\textrm{d}t\,e^{-t^{2}}/\sqrt{\pi}$ is the Gauss error function. Here, the prefactor $N$ ensures that $\lim_{E\to\infty}\mathcal{\overline{N}}(E)=N$. Therefore, we can perform an exact unfolding to each realization of this \textit{ad hoc} integrable system, and use the complete set of unfolded levels, 
 $\{\varepsilon_{i}\}_{i=1}^{N}$ to calculate $\langle P_{k}^{\delta}\rangle$. Note that Eq. \eqref{accnormal} exactly satisfies all the conditions of Theorem 2. 
 
\begin{figure}[h]
\hspace*{-0.3cm}
\includegraphics[width=0.41\textwidth]{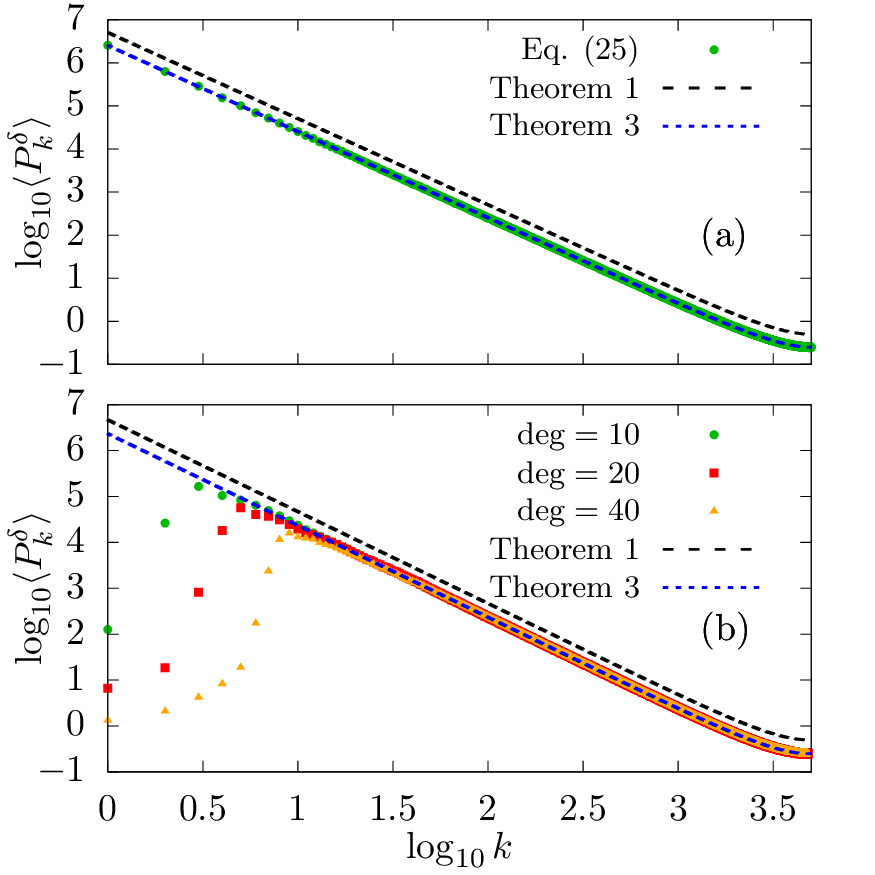}
\caption{Averaged $\delta_{n}$ power-spectrum, $\langle P_{k}^{\delta}\rangle$, for a set of independent, identically distributed, Gaussian eigenlevels $E_{i}\sim\mathcal{G}(0,1)$. Panel (a) shows the result when spectral unfolding is performed exactly by means of the analytic cumulative level function Eq. \eqref{accnormal}; the number of levels in each realization is $N=10^{4}$ and all are used to calculate the results. Panel (b) shows the same but for numerically unfolded levels by means of a polynomial of degree $\textrm{deg}=10,20,40$. Only $0.96N$ levels are kept in this case (see main text). The black, dashed line corresponds to the model of intermediate statistics Eq. \eqref{powerspectrumnounfolding} (see Theorem 1) with $\eta=1$ (Poisson), while the blue, pointed line represents Eq. \eqref{power4} (see Theorem 3) with $\eta=1$ again. }
\label{deltasgaussian}
\end{figure}

 Results are shown in Fig. \ref{deltasgaussian}(a), for the case of $N=10^4$ energy levels, averaged over $10^4$ different realizations. We can see that Theorem 3, Eq. \eqref{power4}, provides a perfect description of the numerical results, in the whole range of frequencies. Even though it has been derived from a slightly different scenario than that coming from Theorem 2, this academic example shows that Theorem 3, Eq. \eqref{power4}, perfectly accounts for the consequences of the unfolding procedure in quantum systems with finite Hilbert spaces. Similar results are expected for physical systems with finite Hilbert spaces for which the smooth part of the cumulative level function can be calculated analytically. Fully connected spin models \cite{Filippone}, like the Lipkin-Meshkov-Glick model \cite{LMG}, for which Eq. \eqref{smoothdensityofstates} can be used in the thermodynamic limit, constitute a paradigmatic example. 
 
In Fig. \ref{deltasgaussian}(b) we show the consequences of performing a numerical unfolding, by means of a polynomial fit so that $\mathcal{\overline{N}}(E)=\sum_{k=0}^{\textrm{deg}}a_{k}E^{k}$, on the same academic example. To perform the calculation, we have removed a total of $0.04N$ levels closest to both spectrum edges;  the remaining $0.96N$ levels $\{\varepsilon_{i}\}_{i}$ are kept to study level statistics (a detailed examination on numerical unfolding will be given below). 
We show the results obtained by means of three different polynomial fits, $\textrm{deg}=10$, $\textrm{deg}=20$, and $\textrm{deg}=40$. Again, Eq. \eqref{power4} gives the correct answer rather transparently, though strong deviations of $\langle P_{k}^{\delta}\rangle$ from the theoretical curve appear at small frequencies. The larger the degree of the polynomial used to fit the smooth part of the cumulative level density, the stronger the deviations and the wider the frequency range at which they appear.
Its origin lies entirely in the polynomial fitting. The resulting curve, $\mathcal{\overline{N}}(E)=\sum_{k=0}^{\textrm{deg}}a_{k}E^{k}$, is the one that minimizes the distance to the exact cumulative level density, ${\mathcal N}(E)$, at the eigenlevels, $\left\{ E_i \right\}$. If the degree of the polynomial is too small, the result provides a poor description of the smooth part of the cumulative level density. On the contrary, if the degree is too large, a part of the fluctuating signal $\widetilde{{\mathcal N}}(E)$ is reproduced by the fit. As a consequence, this part is removed from the $\delta_n$ statistic, and the corresponding frequencies have less power than expected in the final result, $\langle P^{\delta}_k \rangle$. This is precisely what we see in Fig. \ref{deltasgaussian}(b): the larger the degree of the fitting polynomial, the wider the set of frequencies that deviate from the expected result, Eq. \eqref{power4}. As we will see in the next section, the same phenomenon occurs if we rely on polynomial fits to unfold the spectrum of physical systems. A periodogram analysis, like the one proposed in Ref. \cite{Leclair2008}
(which is closely related to the power-spectrum) may be used to determine the best degree for the polynomial fit. 

The next step is to test
the applicability of Theorem 3 to actual quantum many-body systems. Here we rely on a paradigmatic example of a \textit{completely quantum integrable} system, with finite Hilbert space and without a classical analogue, to tackle this task. This is the rational Richardson-Gaudin model on a spin$-1/2$ chain, which is
based on the pairing interaction \cite{Dukelsky2001}. To write down its Hamiltonian we first need to consider the mutually commuting operators
$\left[ R_i, R_j \right]=0$, $\forall i,j\in\{1, \ldots, L\}$, defined by
\begin{equation}
  \label{invariants}
  R_{i}:=K_{i}^{0}+2g\sum_{j\neq
    i}^{L}\frac{1}{\mu_{i}-\mu_{j}}\Bigg\{\frac{1}{2}(K_{i}^{+}K_{j}^{-}+K_{i}^{-}K_{j}^{+})+K_{i}^{0}K_{j}^{0}\Bigg\}.
\end{equation}
They can be understood as acting on a spin-$1/2$ chain with length $L$.
Here, $\mu_{i}$ are $L$ free parameters, $g$ is
the pairing strength and $K^{+}$, $K^{-}$, and
$K^{0}$ are the SU(2) generators, verifying
\begin{equation}
  [K_{\ell}^{0},K_{m}^{+}]=\delta_{\ell
    m}K_{\ell}^{+},\,\,\,\,[K_{\ell}^{+},K_{m}^{-}]=2\delta_{\ell
    m}K_{\ell}^{0}.
\end{equation}

The operators in Eq. \eqref{invariants} allow to construct a fully integrable (i.e., exactly solvable) Hamiltonian,
\begin{equation}\label{rationalhamiltonian}\begin{split}
  &\mathcal{H}_\textrm{Gaudin}:=\sum_{i=1}^{L}\alpha_{i}R_{i}\\&=\sum_{i=1}^{L}\alpha_{i}K_{i}^{0}+g\sum_{j\neq
    i}^{L}\frac{\alpha_{i}-\alpha_{j}}{\mu_{i}-\mu_{j}}\Bigg\{\frac{1}{2}(K_{i}^{+}K_{j}^{-}+K_{i}^{-}K_{j}^{+})+K_{i}^{0}K_{j}^{0}\Bigg\},
\end{split}\end{equation}
where $\alpha_i$ free parameters. This Hamiltonian is quantum integrable because it has $L$ commuting integrals of motion. Moreover, its eigenfunctions can be exactly calculated \cite{Dukelsky2001}. The NNSD has been shown numerically to agree with the Poissonian result $P(s)=e^{-s}$ \cite{Relano2004}.

For our simulations, we work with $L=16$ sites, and have made the
choices $g=1$, $\mu_{i}=i$, and the random variable
$\alpha_{i}\sim\mathcal{G}(0,1)$, $\forall
i\in\{1,\dots,L\}$. Since $[\mathcal{H},K^{0}]=0$, we choose the sector with
$K^{0}=0$, in analogy with the disordered XXZ chain. Thus, the complete set of energies $\{E_{i}\}_{i=1}^{d}$,
with $d=\binom{L}{L/2}=12870$, has been obtained by full matrix
diagonalization. An average over $1000$ realizations has been
performed.

\begin{center}
\begin{figure}
\includegraphics[width=0.5\textwidth]{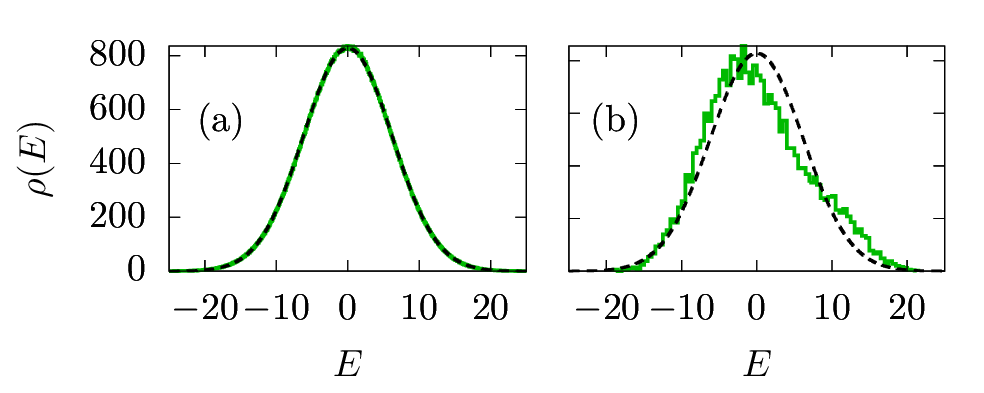} 
\caption{Level density function $\rho(E)$ for the rational Gaudin-Richardson model, Eq. \eqref{rationalhamiltonian} (green), together with a Gaussian fit (black, dashed line), for a number of (a) 1000 realizations and (b) one realization. In both cases, $\rho(E)$ is normalized to the Hilbert space dimension for $L=16$, i.e., $\int _{-\infty}^{\infty}\textrm{d}E\,\rho(E)=12870$.  }
\label{paneldensity}
\end{figure}
\end{center}

\begin{center}
\begin{figure}
\includegraphics[width=0.40\textwidth]{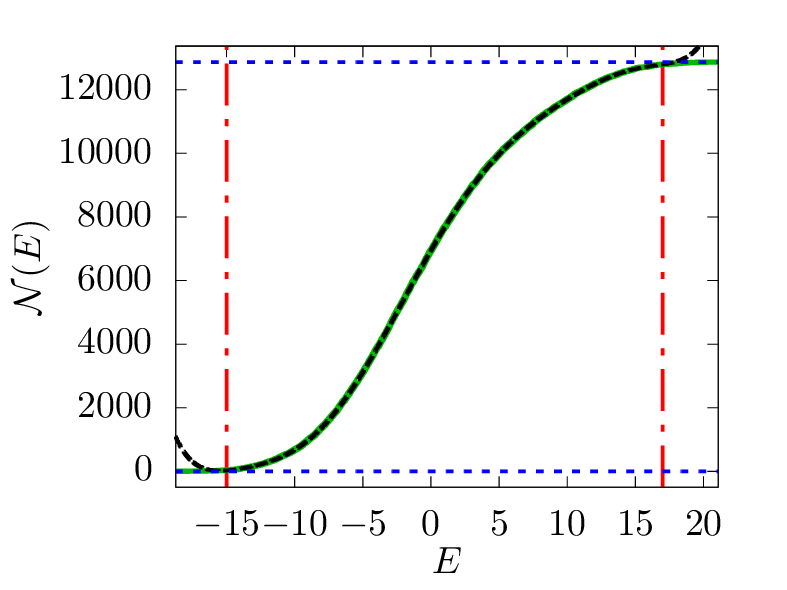} 
\caption{Cumulative level density $\mathcal{N}(E)$ for the rational Gaudin-Richardson model, Eq. \eqref{rationalhamiltonian} (green). The black, dashed line corresponds to the best fit of a generic polynomial of degree 8 to $\mathcal{N}(E)$. Upper and lower blue, pointed lines correspond to the maximum and minimum of $\mathcal{N}(E)$, i.e., to $\mathcal{N}(-\infty)=0$ and $\mathcal{N}(+\infty)=12870$. Strictly, the numerical cumulative function describes the actual $\mathcal{N}(E)$ only within the red, dashed lines, $-15\lesssim E\lesssim 15$.  }
\label{figunfolding}
\end{figure}
\end{center}

In Fig. \ref{paneldensity} we show the level density function $\rho(E)$ resulting from averaging over (a) 1000 realizations and (b) a single realization. In Fig. \ref{paneldensity}(a) we find that $\rho(E)$ very closely resembles a Gaussian, and in fact we have been able to perform a fit to such a curve obtaining excellent results. In Fig. \ref{paneldensity}(b), however, $\rho(E)$ deviates significantly from the (ensemble) numerical fit corresponding to panel (a). In this sense one might say that the Gaudin model is not very ergodic as particular realizations may differ by a substantial amount from the ensemble average, which we believe to be a consequence of the small number of random variables, $L$ (the $\alpha_{i}$), the source of randomness, compared with the total Hilbert space dimension, $L\ll N$.  

Therefore, and even though the smooth part of the density of states is very well described by a Gaussian, it is better to unfold separately each realization of this Hamiltonian by means of a numerical fit. 
To delve into the effects of such a numerical unfolding, in Fig. \ref{figunfolding} we represent the cumulative level density $\mathcal{N}(E)=\int_{-\infty}^{E}\textrm{d}E\,\rho(E)$ corresponding to the level density of Fig. \ref{paneldensity}(b), given by a single realization of Eq. \eqref{rationalhamiltonian}. 
Given the results shown in Fig. \ref{deltasgaussian}(b) for different degrees in the polynomial fit, we have chosen
a polynomial of degree 8, $\sum_{k=0}^{8}a_{k}x^{k}$; the best fit result is shown with black, dashed lines. Blue lines represent $\mathcal{N}(-\infty)=0$ and $\mathcal{N}(+\infty)=12870$. This figure is very transparent about the validity of such common polynomial unfoldings: the polynomial describes very accurately the actual $\mathcal{N}(E)$, but this is true only for values of $E$ \textit{away from the edges}, where the fit clearly cannot account for the real asymptotic behavior of $\mathcal{N}(E)$ (no polynomial is bounded at $\pm\infty$, so this argument is completely general). For this particular case, its range of validity has been schematically indicated by vertical red lines. Thus, it is recommendable to always discard a fraction of levels at the edges before (and, preferably, after as well) spectral unfolding. 
Doing so, the polynomial fit becomes indistinguishable from a non-decreasing function $\overline{{\mathcal N}}(E)$ fulfilling both $\lim_{E \to -\infty} \overline{{\mathcal N}}(E)=0$ and $\lim_{E \to \infty} \overline{{\mathcal N}}(E)=N$, in the energy range in which the unfolded procedure is performed. This means that its consequences are expected to be close to the conditions used to derive Theorem 2, with the possible exception of the lowest frequencies of the statistic $\langle P^{\delta}_k \rangle$, as illustrated by the academic example above.

\begin{center}
\begin{figure}
\includegraphics[width=0.40\textwidth]{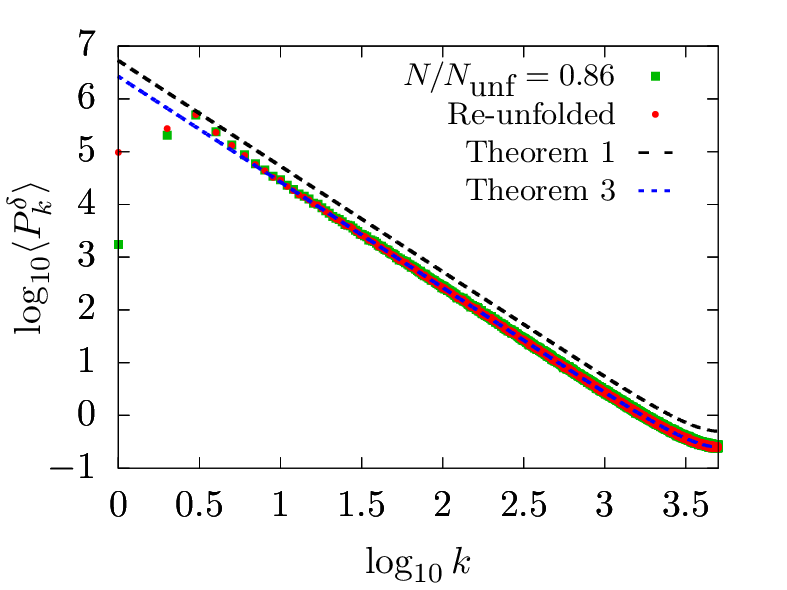}
\caption{Averaged $\delta_{n}$ power-spectrum, $\langle P_{k}^{\delta}\rangle$ for the rational Gaudin-Richardson model, Eq. \eqref{rationalhamiltonian}. The number of sites is $L=16$. Green markers represent the numerical power-spectrum when the usual, polynomial unfolding is performed. Red markers show the numerical power-spectrum when the method of re-unfolding of Theorem 3 is used. In all cases $N/N_{\textrm{unf}}=0.86$. The black, dashed line corresponds to the model of intermediate statistics Eq. \eqref{powerspectrumnounfolding} (see Theorem 1) with $\eta=1$ (Poisson), while the blue, pointed line represents Eq. \eqref{power4} (see Theorem 3) with $\eta=1$ again.}
\label{powgaudingrande}
\end{figure}
\end{center}

 We have fitted the same polynomial to the central $N_{\textrm{unf}}=11870$ levels of the integrable Gaudin model. Then, we have used only a fraction of those levels to study level statistics: these are the central $N=10240$ levels, i.e. $N/N_{\textrm{unf}}\approx 0.86$. The result of this procedure is plotted in Fig. \ref{powgaudingrande} with green squares. We can see a very good agreement between the numerics and the analytical expression provided by Theorem 3, Eq. \eqref{power4}, except for very small frequencies, $k={\mathcal O}(1)$. Hence, our first conclusion is that {\em the consequences of standard unfolding procedures, involving almost the entire spectrum, are well described by Theorem 3} --- but more details will be given later.
 
 We also plot in Fig. \ref{powgaudingrande} the result of {\em re-unfolding} the level spacings, i. e., the result of dividing all of them by the sample estimator of the mean level spacing, $\left< s \right>_{N}$, after the first unfolding is performed. The aim of this re-unfolding is to get closer to the assumptions of Theorem 3. We can see that the results are \textit{indistinguishable from those coming from a standard unfolding procedure}.

\begin{center}
\begin{figure}
\includegraphics[width=0.49\textwidth]{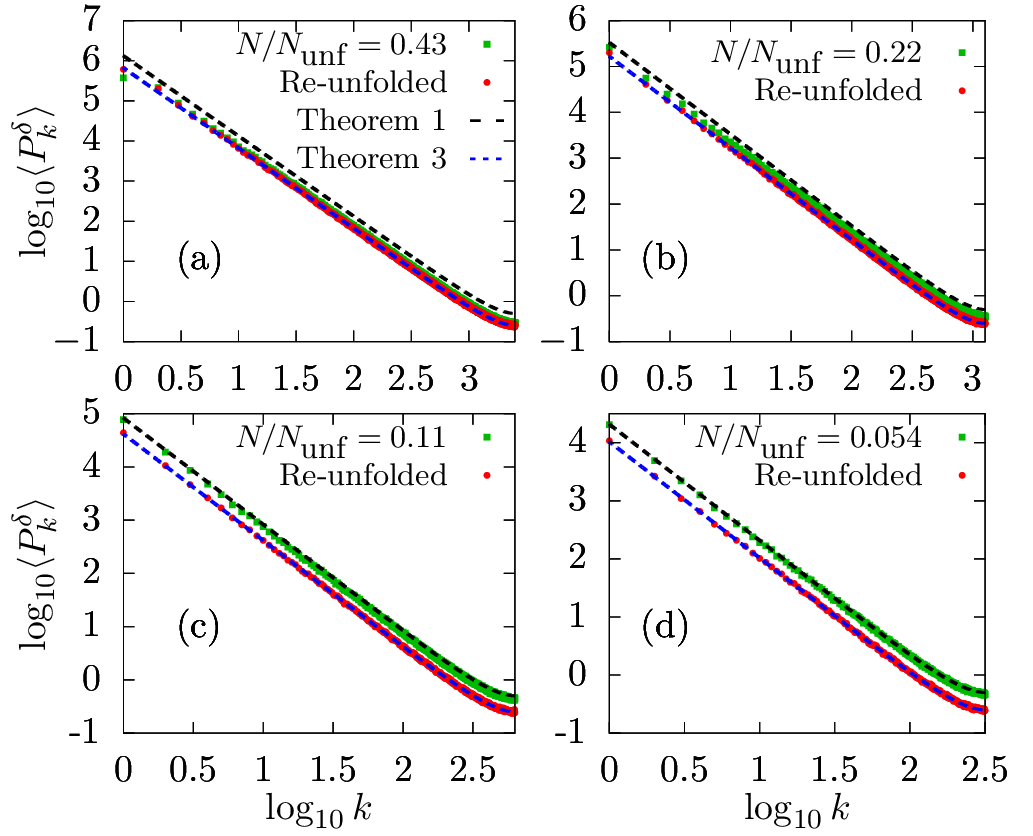} 
\caption{Averaged $\delta_{n}$ power-spectrum, $\langle P_{k}^{\delta}\rangle$ for the rational Gaudin-Richardson model, Eq. \eqref{rationalhamiltonian}. The number of sites is $L=16$. Green markers represent the numerical power-spectrum when the usual, polynomial unfolding is performed with $N_{\textrm{unf}}$ levels and the statistical analysis is done keeping a fraction $N/N_{\textrm{unf}}$. Red markers show the corresponding numerical power-spectrum when the method of re-unfolding is used. The black, dashed line corresponds to the model of intermediate statistics Eq. \eqref{powerspectrumnounfolding} (see Theorem 1) with $\eta=1$ (Poisson), while the blue, pointed line represents Eq. \eqref{power4} (see Theorem 3) with $\eta=1$ again. Panels (a)-(d) show a transition in the fraction of levels used for the statistical analysis $N/N_{\textrm{unf}}$ when $N_{\textrm{unf}}=11870$ is fixed (i.e., $N=5120,2560,1280,640$).}
\label{powgaudintransiton}
\end{figure}
\end{center}

\section{Discussion}\label{discussion}

A first conclusion that can be gathered from Figs. \ref{deltasgaussian} and \ref{powgaudingrande} is that a standard unfolding procedure introduces spurious correlations between energy levels, which mimic the assumptions of Theorems 2 and 3. The first one establishes that the possible values for the $\delta_n$ statistics are highly restricted if the index $n$ is close to the dimension of the system's Hilbert space. However, this restriction becomes much less important for lower values of $n$. For example, if we focus on the center of the spectrum, Theorem 2 establishes that $\delta_{N/2} \leq N/2$. On the other hand, a sequence of independent level spacings with a NNSD given by Eq. \eqref{psgeneralized} entails that $\delta_{N/2}$ is a Gaussian random variable, with zero mean and $\textrm{Var}[\delta_{N/2}]=\sqrt{N/(2\eta)}$, if $N$ is large enough. And the probability for such a random variable to reach, by chance, a value larger than $N/2$ is $\textrm{erfc}(\eta N/\sqrt{2})/2$ [with $\textrm{erfc}(x):=2/\sqrt{\pi}\int_{x}^{\infty}\textrm{d}t\,e^{-t^{2}}$], which is very much negligible for $N \gtrsim 100$ and $\eta\geq1$. Therefore, the restrictions imposed by the unfolding procedure seem irrelevant around the centre of the spectrum, and hence it is logical to wonder what happens if we perform spectral statistics with a \textit{small} number of levels around this spectral region. 

In Fig. \ref{powgaudintransiton} we tackle this task by fixing the number of levels used to perform the unfolding, $N_{\textrm{unf}}$, while keeping only a small and central part of the spectrum to calculate the $\delta_n$ statistic, $N$. The ratio $N/N_{\textrm{unf}}$ is decreased from panel (a), $N/N_{\textrm{unf}}=0.43$, to panel (d), where it is very small, $N/N_{\textrm{unf}}=0.054$. From these results, we can draw the following picture: as the ratio $N/N_{\textrm{unf}}$ decreases, the power-spectrum, $\langle P_{k}^{\delta}\rangle$, obtained by means of the usual polynomial unfolding procedure, drifts from Eq. \eqref{power4} (with $\eta=1$) towards Eq. \eqref{powerspectrumnounfolding} (with $\eta=1$), coinciding with Theorem 1, which does not take into account the spurious correlations introduced by the unfolding procedure.

This is a very reasonable result. As we have advanced above, a small portion of the spectrum around the central level hardly suffers from the restrictions in the $\delta_n$ imposed by Theorem 2. It is worth noting, however, that such a spectral analysis involves just short or medium-range spectral correlations, because the maximum distance between the considered energy levels is much smaller than the dimension of the system's Hilbert space. 
In the case $N\ll N_{\textrm{unf}}$, level spacings behave locally as uncorrelated random variables, leading to Eq. \eqref{powerspectrumnounfolding} rather than to Eq. \eqref{power4}. By contrast, to measure (very) long-range spectral correlations, $N_{\textrm{unf}}$ must be close to the dimension of the Hilbert space, and the ratio $N/N_{\textrm{unf}}$ must be close to unity (and that $N,N_{\textrm{unf}}\gg 1$). This is precisely what we show in Fig. \ref{powgaudingrande}. In this case, the statistical independence is lost between distant levels, and Eq. \eqref{powerspectrumnounfolding}, which is based on independent spacings, no longer affords a good description. 

Besides this important fact, we infer another relevant conclusion from Fig. \ref{powgaudintransiton}. Even in the simple case of a \textit{fully integrable} system such as the Gaudin-Richardson model, performing statistical analysis with Eq. \eqref{powerspectrumnounfolding} as a theoretical result for the power-spectrum becomes very problematic. If it is taken as the theoretical reference, results of Fig. \ref{powgaudintransiton} could be incorrectly used to conclude that this system is {\em not} fully integrable. As small sequences in the centre of the spectrum follow Eq. \eqref{powerspectrumnounfolding} [panel (d)], whereas larger sequences including levels closer to the edges do not [panel (a)], two incorrect conclusions could be gathered: that the dynamical regime is different in different regions of the spectrum, and that the system is not fully integrable because (very) long-range correlations deviate from the Poissonian result. 

In Fig. \ref{powgaudintransiton} we also plot the numerical result that the re-unfolding method provides. We display with red symbols the result of calculating $\langle P_{k}^{\delta}\rangle$ after dividing each sequence of unfolded spacings by $\left< s \right>_{N}$. It is clearly seen that $\langle P_{k}^{\delta}\rangle$ follows Eq. \eqref{power4} with $\eta=1$  \textit{irrespective of the ratio} $N/N_{\textrm{unf}}$. Thus, as opposed to Eq. \eqref{powerspectrumnounfolding}, Eq. \eqref{power4} and the re-unfolding method furnish a good alternative allowing to reach a solid conclusion on the dynamical regime of the quantum many-body system. Hence, as a \textit{practical} byproduct of our theory, we propose to \textit{re-unfold} (that is, to divide every spacing $s_i$ by the corresponding sample estimator of the mean, $s_i \rightarrow s_i/\left<s\right>_{N}$, after the unfolding procedure) any spectrum coming from a quantum system with finite Hilbert space, in order to avoid the problems summarized above.

Putting together Figs. \ref{powgaudingrande} and \ref{powgaudintransiton} we can understand the discrepancy in the repulsion parameter $\eta$ between Figs. \ref{psxxz} and \ref{powxxz}, where we focused on the level statistics of the XXZ chain (away from the ergodic phase).  We remind the reader that the these values, obtained from fitting $P(s)$ and $\langle P_{k}^{\delta}\rangle$ to Eqs. \eqref{psgeneralized} and \eqref{powerspectrumnounfolding}, respectively, did \textit{not} agree by a factor of 2, approximately. As indicated in Theorem 3, Eq. \eqref{power4} takes care of this disagreement. We note that in Fig. \ref{powxxz}, $N/N_{\textrm{unf}}=0.75$, a ratio close to that of Fig. \ref{powgaudingrande} for the Gaudin model. Thus we expect Eq. \eqref{power4} to give a proper characterization of the power-spectrum [instead of Eq. \eqref{powerspectrumnounfolding}, which was used for the fit], and one can approximately obtain the value of $\eta$ in Fig. \ref{psxxz} by dividing the corresponding value in Fig. \ref{powxxz} by 2. We can see that this procedure provides a good description ---although not perfect. It is worth noting that $N_{\textrm{unf}}$ is much smaller than the dimension of the Hilbert space in this case. Therefore, we conclude that the spurious correlations introduced by the unfolding procedure also play a very relevant role when only a part of the spectrum is calculated, as it is frequently done in the study of many-body localization for large chain sizes \cite{shiftinvert}.

\begin{center}
\begin{figure}
\includegraphics[width=0.49\textwidth]{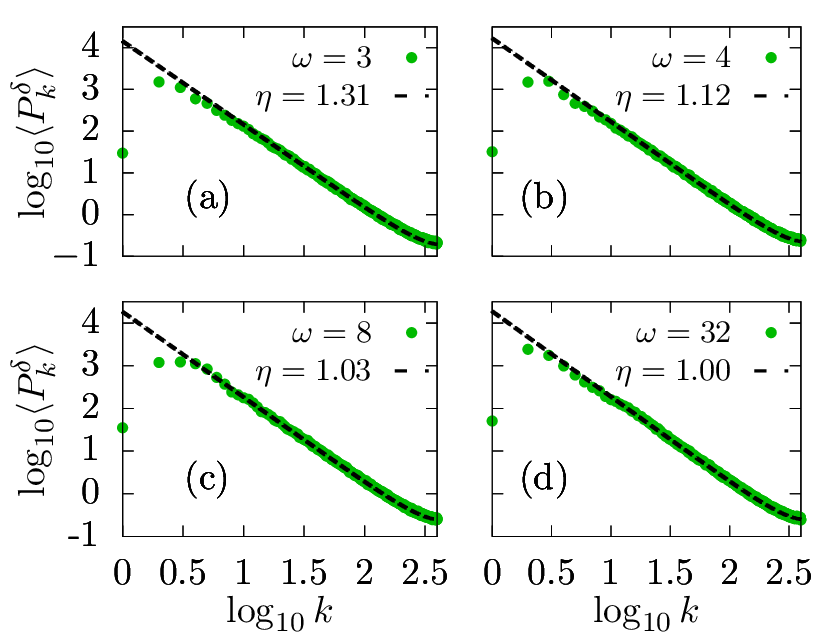}
\caption{Averaged $\delta_{n}$ power-spectrum, $\langle P_{k}^{\delta}\rangle$ for the disordered XXZ Heisenberg chain, Eq. \eqref{modelxxz}. The number of sites is $L=14$. The black, dashed line corresponds to best fit of the model of intermediate statistics Eq. \eqref{power4} to the numerically obtained $\langle P_{k}^{\delta}\rangle$. Here, the method of \textit{re-unfolding} has been used, and in all cases $N/N_{\textrm{unf}}=0.75$. Panels (a)-(d) show the results for disorder strengths $\omega$ and the corresponding value of the repulsion parameter $\eta$ obtained from the fit. Compare this figure with Figs. \ref{psxxz} and \ref{powxxz}. }
\label{pow2xxz}
\end{figure}
\end{center}

Finally, we test the consequences of the re-unfolding procedure. Fig. \ref{pow2xxz} shows the same cases that Fig. \ref{powxxz} (i.e., with $N/N_{\textrm{unf}}=0.75$) but after the energy levels have been re-unfolded, following the method proposed above. We can see that the results for $\eta$, provided by Eq. \eqref{power4}, are now very close to those inferred from short-range spectral statistics, by means of Eq. \eqref{psgeneralized}. This indicates that the re-unfolding procedure is a good tool to study (very) long-range spectral statistics with a proper theoretical reference. Indeed the results shown in Fig. \ref{pow2xxz} allows us to conclude that both short and long-range spectral statistics are compatible with spectra composed of level spacings whose NNSD is given by Eq. \eqref{psgeneralized}.
That is, taking into account the spurious correlations introduced by the unfolding procedure by means of Theorem 3, the results plotted in Fig. \ref{pow2xxz} show that the spectral statistics of the Heisenberg XXZ chain for the considered values of $\omega$ are close to the short-range plasma model proposed in \cite{Bogomolny1999}. 

In closing, we also address one remaining open question, namely the possibility to generalize the main results of this work to infinite-dimensional Hamiltonians when only a finite number of levels are retained for unfolding (even though there are infinite levels). It is our expectation that the conclusions of this paper should also remain valid in that case, but we have always exclusively referred to the finite case because it is in this situation that Theorem 2 can be proved as it stands. If $N$ is allowed to take the `value' $\infty$, then $\max\varepsilon_{n}=\infty$, and thus the bound in Eq. \eqref{deltanmenor1} becomes $\delta_{N-1}\leq\infty$ which, unfortunately, does not really tell much of anything.

\section{Conclusions}\label{conclusions}

In this work we have studied the consequences of the unfolding procedure for quantum systems with finite Hilbert spaces, focusing on many-body quantum systems without a clear semiclassical analogue. 

Taking a celebrated short-range plasma model with independent spacings as a starting point, we have derived the result of Theorem 1: an exact expression for the power-spectrum of the $\delta_{n}$ spectral statistic for a class of intermediate systems that show level repulsion but no statistical correlations as in quantum integrable systems; the fully integrable limit is a particular case of this expression. An investigation of both short- and long-range spectral statistics in the disordered XXZ Heisenberg chain (the paradigmatic model for many-body localization studies) shows that the values of the repulsion parameter $\eta$ obtained from the (short-range) NNSD and the power-spectrum of $\delta_{n}$ (long-range) differ by a factor of approximately two. We argue that this is a consequence of the unfolding procedure, which breaks the global independence of the level spacings from the underlying plasma model. 

Then, we have formulated Theorem 2, showing that the spectrum of \textit{finite} quantum systems cannot be \textit{globally} composed of statistically independent level spacings, not even in the case of fully integrable dynamics. This theorem indicates that the variance of the $\delta_{n}$ statistic calculated from a sequence of truly independent spacings is much larger, and in fact incompatible, with its value after the energy spectrum has been unfolded. It also provides an upper bound for the mean level spacing in an unfolded spectrum: its maximum value is 1 plus a fluctuating term that decreases as $\mathcal{O}(1/N)$. 

To take the consequences of unfolding into account, we have devised a simple model, which we have proposed to call \textit{re-unfolding}, that exactly sets the value of sample estimator of the level spacing to 1. We have re-derived the power-spectrum of $\delta_{n}$ with this model, which is the result of Theorem 3. A numerical study of the rational Gaudin-Richardson model, a fully integrable quantum system without a classical analogue, confirms that this result correctly describes $\langle P_{k}^{\delta}\rangle$ when almost the entire spectrum $\{E_{i}\}_{i=1}^{N}$ is used to estimate the cumulative level function.

Finally, considering that the spurious effects due to the unfolding procedure are expected to be much less important if we consider only a small sequence of levels around the centre of the spectrum, we have numerically studied the consequences of changing the length of this sequence while keeping fixed the number of levels used in the unfolding procedure. Results from the Gaudin model show that as the length of this sequence diminishes (i.e., when only a small number of levels originally used to unfold are kept to actually study level statistics), $\langle P_{k}^{\delta}\rangle$ evolves towards the result of Theorem 1, derived for exactly independent level spacings. On the contrary, when the ratio between the length of the sequence used to calculate $\langle P_{k}^{\delta}\rangle$ and the number of levels used to unfold the original spectrum is close to one, it is Theorem 3 that correctly describes the power-spectrum. As this ratio is necessarily equal to one if we want to study spectral correlations involving all the possible length scales, from consecutive energy levels to levels having $N-2$ other levels in between, being $N$ the dimension of the Hilbert space, this means that, in order to safely study \textit{long-range} spectral statistics and avoid the spurious consequences of unfolding described in this work, it is advantageous to consider the model of re-unfolding here proposed. 

Summarizing, our work shows that, as a consequence of unfolding, the spectrum of any \textit{finite} quantum system shows global (long-range) correlations, even in the case when the dynamics is integrable. We briefly mention that we also expect the unfolding to introduce spurious correlations in quantum chaotic spectra; however, due to their characteristic spectral rigidity, such an effect should play a much less relevant role. Such an investigation would require a separate treatment which lies out of the scope of the present manuscript.

\textit{Note added.---} At a late stage we realized that Eq. \eqref{powerspectrumnounfolding} is in fact formally equivalent to that previously discussed in Ref. \cite{Riser} for uncorrelated spacings, although the connection with the family of intermediate statistics (uncorrelated but with level repulsion)  Eq. \eqref{psgeneralized}, which includes, e.g., the semi-Poisson, had not been made explicit. There, the parameter $\eta$ takes the form of an arbitrary covariance matrix element, not necessarily related  to Eq. \eqref{psgeneralized}, utilized in the study statistics crossovers as in our paper.

\acknowledgments 
The authors thank R. A. Molina for his careful reading and useful suggestions. This work has been financially supported by the Spanish Grants
Nos. FIS2015-63770-P (MINECO/ FEDER) and PGC2018-094180-B-I00
(MCIU/AEI/FEDER, EU).

\appendix

\onecolumngrid

\section{Proof of Theorem 1}\label{prooftheorem1}
From its definition Eq. \eqref{powerspectrum} and applying the linearity of the average operator, $\langle \cdot\rangle$, the averaged power-spectrum of $\delta_{n}$ admits the representation  \begin{equation}\label{pk}
    \langle P_{k}^{\delta}\rangle=\frac{1}{N}\sum_{\ell=1}^{N}\sum_{m=1}^{N}\langle\delta_{\ell}\delta_{m}\rangle e^{i\omega_{k}(\ell-m)}.
\end{equation}
 The starting point for the \textit{correlation factor} is the equation 
\begin{equation}\label{a5}\begin{split}
\langle{\delta}_{\ell}{\delta}_{m}\rangle&=\sum_{i=1}^{\ell}\sum_{j=1}^{m}\big\langle (s_{i}-\langle s_{i}\rangle)(s_{j}-\langle s_{j}\rangle)\big\rangle=\sum_{i=1}^{\ell}\sum_{j=1}^{m}\big(\langle s_{i}s_{j}\rangle -\langle s_{i}\rangle \langle s_{j}\rangle\big). 
\end{split}\end{equation} Each of the terms needed in the above equation are given in Eq. \eqref{sisj}. Thus, performing the double sums yields

\begin{equation}\label{eq11}
    \sum_{i=1}^{\ell}\sum_{j=1}^{m}\langle s_{i}\rangle \langle s_{j}\rangle=\ell m,\hspace{1cm} \sum_{i=1}^{\ell}\sum_{j=1}^{m}\langle s_{i}s_{j}\rangle=\frac{\min\{\ell,m\}}{\eta}+\ell m.
\end{equation} This implies that 

\begin{equation}\begin{split}\label{almost}
    \langle P_{k}^{\delta}\rangle&=\frac{1}{\eta N}\sum_{\ell=1}^{N}\sum_{m=1}^{N}\min\{\ell,m\}e^{i\omega_{k}(\ell-m)}
\end{split}\end{equation}

The remaining double sum can be calculated to give [note that $\omega_{k}N=2\pi k$, and thus $\sin(\omega_{k}N)=0$, $\forall k\in\mathbb{Z}$]
\begin{equation}\label{eq13}\begin{split}
&\sum_{\ell=1}^{N}\sum_{m=1}^{N}\min\{\ell,m\}e^{i\omega_{k}(\ell-m)}=\frac{1}{4\sin^{2}(\omega_{k}/2)}\left[1+2N-\frac{\sin(\omega_{k}N+\omega_{k}/2)}{\sin(\omega_{k}/2)}\right]=\frac{N}{2\sin^{2}(\omega_{k}/2)}.\end{split}\end{equation}

Eqs. \eqref{almost} and \eqref{eq13} imply Eq. \eqref{powerspectrumnounfolding}, and the proof concludes. $\hfill\blacksquare$
\section{Proof of Theorem 3}\label{prooftheorem3}
For the first part of the proof one can proceed in analogy with Theorem 1. However, now the terms in Eq. \eqref{a5} need to be evaluated as they are non-trivial. First, we have
\begin{equation}\label{a6}
\langle \widetilde{s}_{i}\rangle=N\left\langle\frac{s_{i}}{\sum_{j=1}^{N}s_{j}}\right\rangle.
\end{equation} Each ${s}_{i}$ is distributed as in Eq. \eqref{psgeneralized} with probability density function $P_{i}(s_{i}):=\eta^{\eta}s_{i}^{\eta-1}e^{-\eta s_{i}}/\Gamma(\eta)\chi_{[0,+\infty)},$ where $\chi_{A}$ is the characteristic function on $A\subset \mathbb{R}$. Since the set $\{s_{i}\}_{i=1}^{N}$ is of independent random variables, the joint probability density of the $N$-dimensional random variable $\mathbf{s}:=(s_{1},\dots,s_{N})$ factorizes as

\begin{equation}\begin{split}&P(\mathbf{s};\eta):=\prod_{i=1}^{N}P_{i}(s_{i};\eta)=\left(\frac{\eta^{\eta}}{\Gamma(\eta)}\right)^{N}(s_{1}\ldots s_{N})^{\eta-1}e^{-\eta(s_{1}+\ldots+s_{N})}\chi_{[0,+\infty)^{N}}.\end{split}\end{equation} 
Thus, the $i$th nearest neighbor unfolded spacing can be calculated as the $N$-dimensional integral
\begin{equation}
\begin{split}
&\langle \widetilde{s}_{i}\rangle=N\int_{\mathbb{R}^{N}}\prod_{k=1}^{N}\mathrm{d}s_{k}\frac{s_{i}}{s_{1}+\ldots+s_{i}+\ldots+s_{N}}P(\mathbf{s};\eta)=N\left(\frac{\eta^{\eta}}{\Gamma(\eta)}\right)^{N}\int_{0}^{\infty}\prod_{k=1}^{N}\mathrm{d}s_{k}\frac{(s_{1}\ldots s_{N})^{\eta-1}s_{i}^{\eta}}{s_{1}+\ldots+s_{N}}e^{-\eta(s_{1}+\ldots+s_{N})},
\end{split}
\end{equation}
for all $i\in\{1,\ldots,N\}$. To evaluate this integral, we first let $s_{i}=x_{i}^{2}$, so that 

\begin{equation}
    \langle \widetilde{s}_{i}\rangle=2^{N}N\left(\frac{\eta^{\eta}}{\Gamma(\eta)}\right)^{N}\int_{0}^{\infty}\prod_{k=1}^{N}\textrm{d}x_{k} \frac{(x_{1}\ldots x_{N})^{2\eta-1}x_{i}^{2\eta+1}}{x_{1}^{2}+\ldots+x_{N}^{2}}e^{-\eta(x_{1}^{2}+\ldots+x_{N}^{2})}
\end{equation}

Changing to $N$-dimensional hyperspherical coordinates, this integral can be written \begin{equation}\label{16}
\langle\widetilde{s}_{i}\rangle=2^{N}N\left(\frac{\eta^{\eta}}{\Gamma(\eta)}\right)^{N}I_{r}\prod_{k=1}^{N-1}I_{\phi_{k}},\,\,\,\forall i\in\{1,2,\dots,N\},
\end{equation} where \begin{equation}
I_{r}:=\int_{0}^{\infty}\textrm{d}r\,\frac{r^{N-1}r^{2\eta+1}(r^{2\eta-1})^{N-1}}{r^{2}}e^{-\eta r^{2}}=\frac{\Gamma(\eta N)}{2\eta^{\eta N}},
\end{equation}
\begin{equation}
I_{\phi_{1}}:=\int_{0}^{\pi/2}\textrm{d}\phi_{1}\,\cos^{2\eta+1}(\phi_{1})(\sin^{2\eta-1}(\phi_{1}))^{N-1}\sin^{N-2}(\phi_{1})=\frac{\Gamma(1+\eta)\Gamma[\eta(N-1)]}{2\Gamma(1+\eta N)},
\end{equation} and
\begin{equation}
I_{\phi_{i}}:=\int_{0}^{\pi/2}\textrm{d}\phi_{i}\,\cos^{2\eta-1}(\phi_{i})(\sin^{2\eta-1}(\phi_{i}))^{N-i}\sin^{N-i-1}(\phi_{i})=\frac{\Gamma(\eta)\Gamma[\eta(N-i)]}{2\Gamma[\eta(1+N-i)]},\,\,\,i\in\{2,3,\ldots,N-1\}.
\end{equation} After some algebra, Eq. \eqref{16} reduces to the reasonable result that \begin{equation}\label{mean1}
\langle \widetilde{s}_{i}\rangle=1,\,\,\,\forall i\in\{1,\dots,N\}.
\end{equation} Since $\langle\widetilde{s_{j}}\rangle$ is calculated following the exact same procedure, it directly yields the same result. We observe that $\langle \widetilde{s}_{i}\rangle=\langle s_{i}\rangle$, $\forall i\in\{1,\dots, N\}$. This proves the transformation $\widetilde{s}_{i}$ preserves the mean value of the original nearest neighbour spacings, $s_{i}$, as should follow intuitively from its definition.

Next we consider the unfolded nearest neighbour spacing correlation given by 
\begin{equation}\label{a12}
\langle \widetilde{s}_{i}\widetilde{s}_{j}\rangle=N^{2}\left\langle\frac{s_{i}s_{j}}{\sum_{m=1}^{N}\sum_{\ell=1}^{N}s_{m}s_{\ell}}\right\rangle,\,\,\,\forall i,j\in\{1,\dots,N\}.
\end{equation}
For Eq.  \eqref{a12} we can consider two separate cases: $(a)$ $i=j$ and $(b)$ $i\neq j$.

\textit{Case (a).} Consider $i=j$. Then one has \begin{equation}\label{a13}\begin{split}
\langle \widetilde{s}_{i}^{2}\rangle&=N^{2}\left\langle \frac{s_{i}^{2}}{\sum_{m=1}^{N}\sum_{\ell=1}^{N}s_{m}s_{\ell}}\right\rangle=N^{2}\int_{0}^{\infty}\prod_{k=1}^{N}\mathrm{d}s_{k}\frac{s_{i}^{2}}{(s_{1}+\dots+s_{N})^{2}}P(\mathbf{s};\eta)\\&=N^{2}\left(\frac{\eta^{\eta}}{\Gamma(\eta)}\right)^{N}\int_{0}^{\infty}\prod_{k=1}^{N}\textrm{d}s_{k}\,\frac{(s_{1}\ldots s_{N})^{\eta-1}s_{i}^{\eta+1}}{(s_{1}+\ldots+s_{N})^{2}}e^{-\eta(s_{1}+\ldots+s_{N})},
\end{split}\end{equation} 
for all $i\in\{1,2,\ldots,N\}$. To evaluate the above integral, we first make the change of variables $s_{i}=x_{i}^{2}$ so that 

\begin{equation}
    \langle \widetilde{s}_{i}^{2}\rangle=2^{N}N^{2}\left(\frac{\eta^{\eta}}{\Gamma(\eta)}\right)^{N}\int_{0}^{\infty}\prod_{k=1}^{N}\textrm{d}x_{k}\,\frac{(x_{2}\ldots x_{N})^{2\eta-1}x_{1}^{2\eta+3}}{(x_{1}^{2}+\ldots+x_{N}^{2})^{2}}e^{-\eta(x_{1}^{2}+\ldots+x_{N}^{2})}.
\end{equation}

Now we can further change variables to $N$-dimensional hyperspherical coordinates as before, which allows us to rewrite the above equation in the form 

\begin{equation}
    \langle\widetilde{s}_{i}^{2}\rangle=2^{N}N^{2}\left(\frac{\eta^{\eta}}{\Gamma(\eta)}\right)^{N}\mathcal{I}_{r}\prod_{k=1}^{N-1}\mathcal{I}_{\phi_{k}}.
\end{equation}
The quantities $\mathcal{I}_{r}$ and $\mathcal{I}_{\phi_{k}}$ are the radial and the $N-1$ angular integrals over $\mathbb{R}^{N}$, respectively. They are given by 

\begin{equation} \mathcal{I}_{r}:=\int_{0}^{\infty}\textrm{d}r\,\frac{r^{N-1}r^{2\eta+3}}{r^{4}}(r^{N-1})^{2\eta-1}e^{-\eta r^{2}}=\frac{\Gamma(\eta N)}{2\eta^{\eta N}},\end{equation}
\begin{equation}\begin{split}
\mathcal{I}_{\phi_{1}}&:=\int_{0}^{\pi/2}\textrm{d}\phi_{1}\,\sin^{N-1}(\phi_{1})\cos^{2\eta+3}(\phi_{1})\sin^{(N-1)(2\eta-1)}(\phi_{1})=\frac{\Gamma(\eta+2)\Gamma[\eta(N-1)]}{2\Gamma(2+\eta N)}, \end{split}\end{equation}
and 

\begin{equation}\begin{split}
\mathcal{I}_{\phi_{i}}&:=\int_{0}^{\pi/2}\textrm{d}\phi_{i}\,\sin^{N-i-1}(\phi_{i})\left[\cos(\phi_{i})\sin^{N-i}(\phi_{i})\right]^{2\eta-1}=\frac{\Gamma(\eta)\Gamma[\eta(N-i)]}{2\Gamma[\eta(N-i+1)]},\,\,\,\forall i\in\{2,\ldots,N-1\}. \end{split}\end{equation}
After some algebra one can simplify the result to obtain \begin{equation}\langle \widetilde{s}_{i}^{2}\rangle=\frac{(\eta+1)N}{1+\eta N},\,\,\,\forall i\in\{1,\dots,N\}.
\end{equation}
\textit{Case (b).} Consider now $i\neq j$. Following the same steps we obtain instead \begin{equation}\label{a14}\begin{split}
&\langle \widetilde{s}_{i}\widetilde{s}_{j}\rangle=N^{2}\left\langle \frac{s_{i}s_{j}}{\sum_{m=1}^{N}\sum_{\ell=1}^{N}s_{m}s_{\ell}}\right\rangle=N^{2}\int_{0}^{\infty}\prod_{k=1}^{N}\mathrm{d}s_{k}\frac{s_{i}s_{j}}{(s_{1}+\dots+s_{N})^{2}}P(\mathbf{s};\eta)\\&=N^{2}\left(\frac{\eta^{\eta}}{\Gamma(\eta)}\right)^{N}\int_{0}^{\infty}\prod_{k=1}^{N}\textrm{d}s_{k}\,\frac{(s_{1}\ldots s_{N})^{\eta-1}s_{i}^{\eta}s_{j}^{\eta}}{(s_{1}+\ldots+s_{N})^{2}}e^{-\eta(s_{1}+\ldots+s_{N})},
\end{split}\end{equation} 
for all $i,j\in\{1,2,\ldots,N\}$, $i\neq j$. Making the same variable changes as in case $(a)$, this integral can be cast in the form 

\begin{equation}
    \langle \widetilde{s}_{i}\widetilde{s}_{j}\rangle=2^{N}N^{2}\left(\frac{\eta^{\eta}}{\Gamma(\eta)}\right)^{N}\mathfrak{I}_{r}\prod_{k=1}^{N-1}\mathfrak{I}_{\phi_{k}}.
\end{equation}
The radial integral yields the same result as before,

\begin{equation}
    \mathfrak{I}_{r}:=\int_{0}^{\infty}\textrm{d}r\,\frac{r^{N-1}(r^{N-2})^{2\eta-1}}{r^{4}}r^{4\eta+2}e^{-\eta r^{2}}=\frac{\Gamma(\eta N)}{2\eta^{\eta N}}
\end{equation}
while the $N-1$ angular integrals are now

\begin{equation}\begin{split}
\mathfrak{I}_{\phi_{1}}&:=\int_{0}^{\pi/2}\textrm{d}\phi_{1}\,\sin^{N-2}(\phi_{1})\cos^{2\eta+1}(\phi_{1})\sin^{2\eta+1}(\phi_{1})(\sin^{N-2}(\phi_{1}))^{2\eta-1}=\frac{\Gamma(\eta+1)\Gamma[\eta(N-1)+1]}{2\Gamma(2+\eta N)},
\end{split}\end{equation}

\begin{equation}\begin{split}
\mathfrak{I}_{\phi_{2}}&:=\int_{0}^{\pi/2}\textrm{d}\phi_{2}\,\sin^{N-3}(\phi_{2})\cos^{2\eta+1}(\phi_{2})(\sin^{N-2}(\phi_{2}))^{2\eta-1}=\frac{\Gamma(\eta+1)\Gamma[\eta(N-2)]}{2\Gamma[\eta(N-1)+1]},
\end{split}\end{equation}

and 

\begin{equation}\begin{split}
\mathfrak{I}_{\phi_{i}}&:=\int_{0}^{\pi/2}\textrm{d}\phi_{i}\,\sin^{N-i-1}(\phi_{i})[\cos(\phi_{i})\sin^{N-i}(\phi_{i})]^{2\eta-1}=\frac{\Gamma(\eta)\Gamma[\eta(N-i)]}{2\Gamma[\eta(N-i+1)]},\,\,\,i\in\{3,4,\ldots,N-1\}.
\end{split}\end{equation}

Simplifying now yields \begin{equation}\langle \widetilde{s}_{i}\widetilde{s}_{j}\rangle=\frac{\eta N}{1+\eta N},\,\,\,\forall i,j\in\{1,\dots,N\},\,i\neq j.\end{equation} Both cases $(a)$ and $(b)$ can be expressed compactly, producing the full correlator
\begin{equation}\label{a15}
\langle \widetilde{s}_{i}\widetilde{s}_{j}\rangle=\frac{N(\eta+\delta_{ij})}{\eta N+1},\,\,\,\forall i,j\in\{1,\dots,N\}.
\end{equation} 

Plugging Eqs. (\ref{a6}) and (\ref{a15}) into Eq. \eqref{a5}, we obtain 

\begin{equation}\label{a16}\begin{split}
\langle {\delta}_{\ell}{\delta}_{m}\rangle&=\sum_{i=1}^{\ell}\sum_{j=1}^{m}\left[\frac{N(\eta+\delta_{ij})}{1+\eta N}-1\right]=\frac{1}{\eta N+1}\left(N\min\{\ell,m\}-\ell m\right).
\end{split}
\end{equation}
The power-spectrum is now rewritten 

\begin{equation}\label{35}
\langle P_{k}^{\delta}\rangle=\frac{1}{N(\eta N+1)}\sum_{\ell=1}^{N}\sum_{m=1}^{N}\left[N\min(\ell,m)-\ell m\right]e^{i\omega_{k}(\ell-m)}.
\end{equation}

Proceeding requires performing the two double sums that appear in Eq. \eqref{35}. The first one is Eq. \eqref{eq13}. The second one is found to be

\begin{equation}\label{sum2}\begin{split}
\sum_{\ell=1}^{N}\sum_{m=1}^{N}\ell m e^{i\omega_{k}(\ell-m)}&=\frac{1}{8\sin^{4}(\omega_{k}/2)}\Bigg\{1+N+N^{2}-(1+N)[\cos(\omega_{k}N)+N\cos(\omega_{k})]+N\cos(\omega_{k}+\omega_{k}N)\Bigg\}\\&=\frac{N^{2}}{8\sin^{4}(\omega_{k}/2)}\left[1-\cos(\omega_{k})\right]=\frac{N^{2}}{4\sin^{2}(\omega_{k}/2)}.\end{split}\end{equation} In the above equations, use of the facts that $\cos(\omega_{k}N)=1$ and $\sin(\omega_{k}N)=0$, $\forall k\in\mathbb{Z}$, has been made. Finally, inserting Eqs. \eqref{eq13} and \eqref{sum2} into Eq. \eqref{35}, one obtains the desired \textit{exact} representation Eq. \eqref{power4}.

This proves the theorem. $\hfill\blacksquare$
\twocolumngrid


\begin{thebibliography}{100}
\bibitem{exp1} Given two random variables $X$ and $Y=\alpha X$, $\alpha\in\mathbb{R}$, with probability density functions $f_{X}(x)$ and $f_{Y}(y)$, one has that $f_{Y}(y)=\frac{1}{|\alpha|}f_{X}\left(x=\frac{y}{\alpha}\right)$.
\bibitem{exp2} To simplify the notation of the calculations, here and in Theorem 3 we consider $N+1$ levels so that there are exactly $N$ level spacings.
\bibitem{exp3} Recall that, by Eq. \eqref{deltan}, the last value at which  $\delta_{n}$ can be evaluated in a spectrum of $N$ levels is $N-1$.
\end{thebibliography}
\end{document}